\documentclass[a4paper,fleqn,usenatbib]{mnras}

\usepackage[T1]{fontenc}
\usepackage{ae,aecompl}

\usepackage{graphicx}	
\usepackage{amsmath}	
\usepackage{amssymb}	
\usepackage{booktabs}

\newcommand{\mAA}{\textup{\AA}}
\newcommand{\xmm}{\rm{XMM-Newton}}
\newcommand{\xspec}{{\sc Xspec}}

\newcommand{\Lhard}{$L_{\text{2--10 keV}}$}
\newcommand{\Lsoft}{$L_{\text{0.5--2 keV}}$}

\newcommand{\bknpow}{{\sc wabs*zwabs*bknpower}}
\newcommand{\hardpl}{{\sc wabs*zwabs*powerlaw}}
\newcommand{\bb}{{\sc wabs*(zbbody+powerlaw)}}

\newcommand{\alphaox}{$\alpha_{OX}$}
\newcommand{\alphaio}{$\alpha_{IO}$}

\newcommand{\cMdot}{$\dot{\mathcal{M}}$}
\newcommand{\mdot}{$\dot{m}$}
\newcommand{\cMdotcrit}{$\dot{\mathcal{M}}_{min}$}

\newcommand{\Lbol}{$L_{\text{AGN}}$}
\newcommand{\Lopt}{$L_{5100\mAA}$}

\title[Unveiling slim AD]{Unveiling slim accretion disc in AGN through X-ray and Infrared observations}

\author[Castell\'o-Mor et al.]{
N\'uria Castell\'o-Mor,$^{1}$\thanks{E-mail: nuria@wise.tau.ac.il (TAU)}
Shai Kaspi,$^{1,2}$
Hagai Netzer,$^{1}$
Pu Du,$^{3}$
Chen Hu,$^{3}$
Luis C. Ho,$^{4,5}$
\newauthor Jin-Ming Bai,$^{6}$
Wei-Hao Bian,$^{7}$
Ye-Fei Yuan,$^{8}$
Jian-Min Wang$^{3,9}$
\\
$^{1}$School of Physics and Astronomy, Tel Aviv University, Tel Aviv 69978, Israel\\
$^{2}$Wise Observatory, School of Physics and Astronomy, Tel Aviv University, Tel Aviv 69978, Israel\\
$^{3}$Key Laboratory for Particle Astrophysics, Institute of High Energy Physics, Chinese Academy of Sciences,
19B Yuquan Road,\\ Beijing 100049, China\\
$^{4}$Kavli Institute for Astronomy and Astrophysics, Peking University, Beijing 100871, China\\
$^{5}$Department of Astronomy, School of Physics, Peking University, Beijing 100871, China\\
$^{6}$Yunnan Observatories, Chinese Academy of Sciences, Kunming 650011, China\\
$^{7}$Physics Department, Nanjing Normal University, Nanjing 210097, China\\
$^{8}$Department of Astronomy, University of Science and Technology of China, Hefei 230026, China\\
$^{9}$National Astronomical Observatories of China, Chinese Academy of Sciences, 20A Datun Road, Beijing 100020, China\\
}

\date{Accepted XXX. Received YYY; in original form ZZZ}

\pubyear{2015}

\begin{document}
\label{firstpage}
\pagerange{\pageref{firstpage}--\pageref{lastpage}}
\maketitle

\begin{abstract}
    In this work, which is a continuation of \citet{Castello2016}, we present
    new X-ray and infrared (IR) data for a sample of
    active galactic nuclei (AGN) covering a wide range in Eddington ratio over a
    small luminosity range.
    In particular, we rigorously explore the dependence of the optical-to-X-ray
    spectral index \alphaox{} and the IR-to-optical spectral index on the dimensionless
    accretion rate, \cMdot{}$=\dot{m}/\eta$ where \mdot$=L_{AGN}/L_{Edd}$ and
    $\eta$ is the mass-to-radiation conversion efficiency, in low and high accretion
    rate sources. We find that the SED of the faster accreting sources are surprisingly
    similar to those from the comparison sample of sources with lower accretion rate.
    In particular:
    {\sc i)} the optical-to-UV AGN SED of slow and fast accreting AGN can be
    fitted with thin AD models.
    {\sc ii)} The value of \alphaox{} is very similar in slow and fast accreting
    systems up to a dimensionless accretion rate \cMdot$_{c}\sim$10. We only find a
    correlation between \alphaox{} and \cMdot{} for sources with
    \cMdot$>$\cMdot$_{c}$. In such cases, the faster accreting sources appear to have
    systematically larger \alphaox{} values.
    {\sc iii)} We also find that the torus in the faster accreting systems seems to be less
    efficient in reprocessing the primary AGN radiation having lower IR-to-optical spectral
    slopes.\\
    These findings, failing to recover the predicted differences between the SEDs of slim
    and thin ADs within the observed spectral window, suggest that additional physical
    processes  or very special geometry act to reduce the extreme UV radiation in fast accreting
    AGN. This may be related to photon trapping, strong winds, and perhaps other
    yet unknown physical processes.
\end{abstract}

\begin{keywords}
galaxy -- quasar -- Seyfert
\end{keywords}


\section{Introduction}

This work is a continuation of our previous paper \citep{Castello2016} where we
studied the properties of thin and slim accretion discs (AD) in active galactic
nuclei (AGN). Theory postulated profound differences between the typical
spectral energy distribution (SED) of slow and fast accreting black holes (BHs)
\citep{Wang1999,Mineshige2000}. In slow accreting systems, the primary emission
of the AGN is well determined by the standard thin AD
theory \citep[][hereafter SS73]{Shakura1973}, where the total luminosity emitted
by the AGN, $L_{AGN}$, is set by the accretion rate (\mdot$=L_{AGN}/L_{Edd}$)
and the mass-to-radiation conversion efficiency, $\eta$, which, in itself, depends
on the BH spin. The direct integration of the disc SED is not practical in almost
all cases, because much of the radiation is emitted beyond the Lyman limit at
912$\,$\AA{}. To avoid the
dependence on the BH spin, which sets the total emitted radiation in
thin ADs, a dimensionless accretion rate \cMdot{} can be introduced
\begin{equation}
  \label{calMdot}
    \dot{\mathcal{M}}\equiv\dot{m}/\eta \simeq 20.1 \left(\frac{l_{44}}{\cos{i}}\right)^{3/2} M_{7}^{-2}
\end{equation}
where $M_{7}=M_{BH}/10^7 M_{\odot}$, $i$ the inclination angle and $l_{44}$ the
luminosity at 5100\AA{} in units of $10^{44}$~erg/s \citep[see][]{Frank2002,Netzer2013,DuPu2015}.
Thus, for \mdot$=$1, a non-rotating BH results in \cMdot$=17.54$ and a
maximally prograde rotating BH gives \cMdot$=$3.1.

The BH governs a wide range of observed AGN properties such as spectral
slopes and other emission properties. It also affects the geometry of the AD.
Beyond a critical mass accretion rate \cMdot$_{c}$, energy transport by
advection dominates over radiative in the inner parts of the system and
the disc is thought to become geometrically thick, or slim \citep[``slim
accretion disc'',][]{Abramowicz1988, Laor1989,
Wang1999, Sadowski2011, Kawakatu2011}. In such cases, the standard AD model
breaks down and processes that are not important in geometrically thin ADs must
be taken into account. Theoretical works \citep{Wang1999,Mineshige2000} suggest
that in the slim disc regime, photon trapping, followed by radial advection, reduces
the mass-to-radiation efficiency leading to saturated luminosity. In particular,
the physics related to the innermost stable orbit changes dramatically and the
BH spin is not affecting the radiation efficiency factor $\eta$. While these
general ideas have been suggested in numerous slim disc calculations, they are
not confirmed by recent numerical simulations \citep[][and references
therein]{Sadowski2016} which indicate that radial advection is far less important
than estimated by the earlier models. They also suggest that much of the
released energy can be via disc winds and the fraction of emitted radiation,
its SED and angular pattern, are highly uncertain.
There are also disagreements among the various simulations \citep[see
e.g.][]{Jiang2014} indicating how immature this field is.
Other works, such us \citep{Collin2004}, are focused on the
intrinsic characteristics of the general population.

Slim AD systems are thought to have extreme X-ray properties. The reason is not fully
understood with the best explanations so far relating this to
the conditions in the X-ray emitting corona via Comptonization, photon trapping, magnetorotational
instabilities, and extreme radiation pressure force \citep[see review in ][]{Wang2013}.

When comparing thin to slim accretion discs it is important to note that in both
systems, the observed 5100\AA{} emission ($l_{44}$) is thought to originate at
large enough radii and thus is less influenced by the radial motion of the
accretion flow compared with the regions closer in that emit the shorter
wavelength photons.
At these large radii Eq.~\ref{calMdot} is a reliable way to determine
the normalized mass accretion rate of slim AD systems.

Following earlier works on slim ADs \citep[e.g.][and references
therein]{DuPu2015}, we classified candidates to host slim discs as those objects with
\mdot$=\eta$\cMdot{}$\geq$0.1. This is based on both the thickening of the disc,
as well as, general considerations about the onset of radial advection in the
inner parts of the disc.
Since we cannot observe the entire SED, we have no direct way to measure
\mdot{}. To be conservative, we chose the lowest possible efficiency,
$\eta=0.038$ (retrograde accretion) to set a lower limit on the normalized
accretion rate, \cMdot$=$2.63. For simplicity we change this rate slightly and
assume that \cMdot$_{min}=$3 is the requiered minimum normalzied accretion rate
for a slim AD candidate.

In 2013 we started a large observational project aimed at understanding the faster accreting AGN in the local Universe.
The first part of our project involved accurate determination of BH masses through reverberation mapping
\citep[RM,][]{DuPu2014,DuPu2015,Wang2014,DuPu2015}, and hence its mass accretion rate.
We found that fast accreting systems do not follow the well-known correlation between the typical broad line
region (BLR) size, $R_{BLR}$, and the monochromatic luminosity at 5000\,\AA{}, \Lopt, having lower BH masses
for a given luminosity. We interpret this behaviour as a change in the geometry of the AD from thin to slim.
Furthermore, \citet{Castello2016} (hereafter paper~{\sc I}) compared IR-optical-UV-X-ray SEDs of 16 fast accreting systems
(\cMdot$\geq$\cMdot$_{min}$) with RM-based BH masses, to a sample of 13 RM-mapped AGN with low accretion
rates (\cMdot$<$\cMdot$_{min}$). While AD theory predicts slim AD systems to have much higher
UV luminosity, observations falsify this prediction. In paper~I, we see no evidence for extremely luminous ionizing
continua, and no unusual torus emission (albeit from large aperture WISE data), in those sources expected
to be powered by slim ADs.
While the spectral differences between fast and slow accretors are indirect, since the
observations do not reach far enough into the UV where slim and thin AD emit most of their radiation, it is
hard to believe that the increase in the total disc luminosity expected in those AGN that are
powered by slim ADs, is exactly compensated for by a changing geometry of the
tours.

In this paper we expand the work to include larger AGN samples and new X-ray and IR data. The goal is
to understand better the role of accretion rate in determining the physics of
high Eddington ratio sources over a well defined, limited luminosity range and to critically test various
suggestions about the differences between thin and slim ADs.
In particular we want to compare the role of accretion rate on the 2~keV and 5$\mu$m emission
in low and high Eddington ratio sources.
Note that the limited luminosity range refers to the range in \Lopt{}, not in
the total luminosity emitted by the AGN, \Lbol{}. The conversion to total
luminosity involves the BH mass and spin and hence the limit on \Lbol{} is not
constrained so well.
Throughout the paper we make a clear distinction
between fast accreting AGN and narrow-line Seyfert 1 (NLS1s) since many objects in the latter group are
not necessarily high Eddington ratio sources.

The structure of the paper is as follows: In Section 2, we describe the sample used in this study, how
we define low and high accretion rate AGN, and how BH mass and dimensionless
accretion rate are estimated.
In Section~\ref{DataAnalysis}, we describe how the spectral properties of the AGN samples were deduced.
The correlation analysis between \alphaox{} and the new IR-to-optical spectral
slope on the dimensionless accretion rate is presented in Section~\ref{Results}. Finally,
in Section~\ref{Conclusions} we summarize our findings. In Appendix~\ref{appendix}, we present
the X-ray spectral analysis for new RM-selected X-ray sources
with the highest Eddington ratio up to now.

\section{The Sample}
\label{TheSample}

The AGN employed in the present analysis are obtained from two different samples with the objective
of spanning a large range of accretion rates.
From the 29 AGN presented in paper~{\sc I}, we select all the sources that have X-ray measurements.
We then selected 15 out of the 29 AGN that have been observed by \xmm{} in the past and four additional
sources (J075101, J080101, J081441, and J100055) that have been recently observed
by \xmm{} especially for this project (PI: Shai Kaspi). These 19/29 objects (of which 11 have \cMdot$<$3)
belong to a unique sample via RM \citep{DuPu2014,DuPu2015} and hence their BH mass was measured.
As of early 2015, the sample include
basically all the potential slim disc candidates which are bright enough in the optical/UV to have direct
mass measurements \citep[][]{DuPu2015}.

A complementary sample of 36 AGN were selected from the bright soft X-ray selected AGN sample of \citet{Grupe2010}.
Because the aim of our work is to investigate the connection of \alphaox{} with
the dimensionless accretion rate, this
sample was selected to overlap in optical luminosity (defined as $\lambda L_{\lambda}$ at the rest-frame
wavelength 5100\AA{}) the luminosity of the RM-selected AGN sample. We note that
this sample represents the largest number of AGN observed to study contemporaneous optical-to-X-ray observations
within a well-defined luminosity range.
\citet{Grupe2010} make the common assumption that the black hole mass scaling relation derived from RM-selected
AGN sample is applicable to all AGN.
However, \citet{DuPu2015} found that fast accreting AGN do not follow the well-known correlation between $R_{BLR}$
and \Lopt.
This result was confirmed by the more recent study of \citet{Du2016} where
several additional objects with very high Eddington ratio are are included. In this work, the BH mass of the \citet{Grupe2010} AGN (hereafter
\emph{soft-X-ray-selected} AGN group) were estimated assuming
that the $R_{BLR}$--\Lopt{} relationship is in fact a function of the mass accretion rate
(see Section~\ref{BHmassAccretionSection}).

\subsection{BH mass and dimensionless accretion rate}
\label{BHmassAccretionSection}

Direct determination of BH mass
through RM (which is the most accurate technique) still only exist for a small fraction of
all AGN ($\sim$60), and bolometric
luminosities are subject to large uncertainties about the extreme-UV part of the SED which is not directly observable.
In the present paper we follow the \citet{DuPu2015} method to determine a dimensionless accretion rate,
\cMdot{}, which is independent of the mass-to-radiation, spin-dependent conversion efficiency
$\eta$ (see Eq.~\ref{calMdot}).
The equation we use is based on the standard way to calculate the accretion rate
in the SS73 thin AD model \citep[for more explanations see][]{Frank2002,Netzer2013}.
We assumed an inclination angle of $\cos i=0.75$ which corresponds an inclination typical of type-I AGN that
cannot be observed from a much larger inclination angle due to obscuration by the central torus.

For the RM-selected AGN group, the BH mass are taken from the RM papers and are based on measurements of the
$FWHM(H\beta)$ method \citep[see][]{DuPu2015}. For the
soft-X-ray-selected AGN group, for which the BH mass is unknown, we need to distinguish between low
and high mass accretion rate sources since the two groups are known to have different $R_{BLR}$~--~\Lopt{}
relationships. We therefore use the following two-step approach. We compute $M_{BH}$ based on the relationship
found for slow accreting AGN \citep{Kaspi2000,Bentz2009,Bentz2013,DuPu2015}. In this case
$R_{BLR}=\mathcal{K} l_{44}^{\alpha}$ and the BH mass is
\begin{equation}
    \label{eq.Mbh}
    M_{BH} = 1.4625\times10^5 f_{BLR} \left(\frac{FWHM}{10^3\text{km/s}}\right)^2 \mathcal{K} l_{44}^{\alpha}
\end{equation}
where FWHM is the full-width
at half-maximum intensity of H$\beta$ and $f_{BLR}$ is a factor that includes information about
the geometry and kinematics of the BLR gas. We assumed $f_{BLR}=1$ in agreement with \citet{DuPu2015} and
the recent work of \citet{Woo2015}.  The slope $\alpha$ and the scaling factor $\mathcal{K}$ differ
slightly from one study to the next.
In the first stage we estimate the BH mass and \cMdot{} of the soft-X-ray-selected AGN using the
$R_{BLR}-l_{44}$ correlation given by \citet[][$\mathcal{K}=33.65$ and $\alpha=0.53$]{Bentz2013} which assumes
no dependence of $R_{BLR}$ on the accretion rate.
We then use equation~\ref{calMdot} to obtain a rough estimate of \cMdot{} and thus determine
whether a dependece on
accretion rate is necessary.
As explained, the fiducial accretion rate we use to set the
boundary between the two groups is \cMdot$_{min}=$3, which corresponds to the minimum
possible accretion rate to be selected as an slim AD system (see Introduction).
According to \citet{Du2016} for such objects
\begin{equation}
    \label{eq.cMdotc}
    R_{BLR}=\mathcal{K} l_{44}^{\alpha} \text{  min}\left[1, \left(\frac{\dot{\mathcal{M}}}{\dot{\mathcal{M}}_{c}}\right)^{-\gamma} \right]
\end{equation}
Unfortunately, this expression does not allow simple iteration on the BH mass since
starting with the mass obtained in this way (i.e. the parameters used for the slow
accreting systems) and calculating new $M_{BH}$ and \cMdot{} leads to divergence.
We therefore followed a different procedure based on a different equation of the
type $R_{BLR}=\mathcal{K}l_{44}^{\alpha}$, which is not meant to fit all high accretion
rate sources, but rather to find a good approximation, over a limited luminosity range, that is changing
continuously with \Lopt{} and is consistent with Eq.~\ref{eq.cMdotc} for the
sources with \cMdot{}$\geq$\cMdotcrit{}. The parameters that fit these requirements very well are
$\mathcal{K}=14.96$ and $\alpha=0.14$.

\begin{figure}
    \centering
    \includegraphics[width=0.8\linewidth]{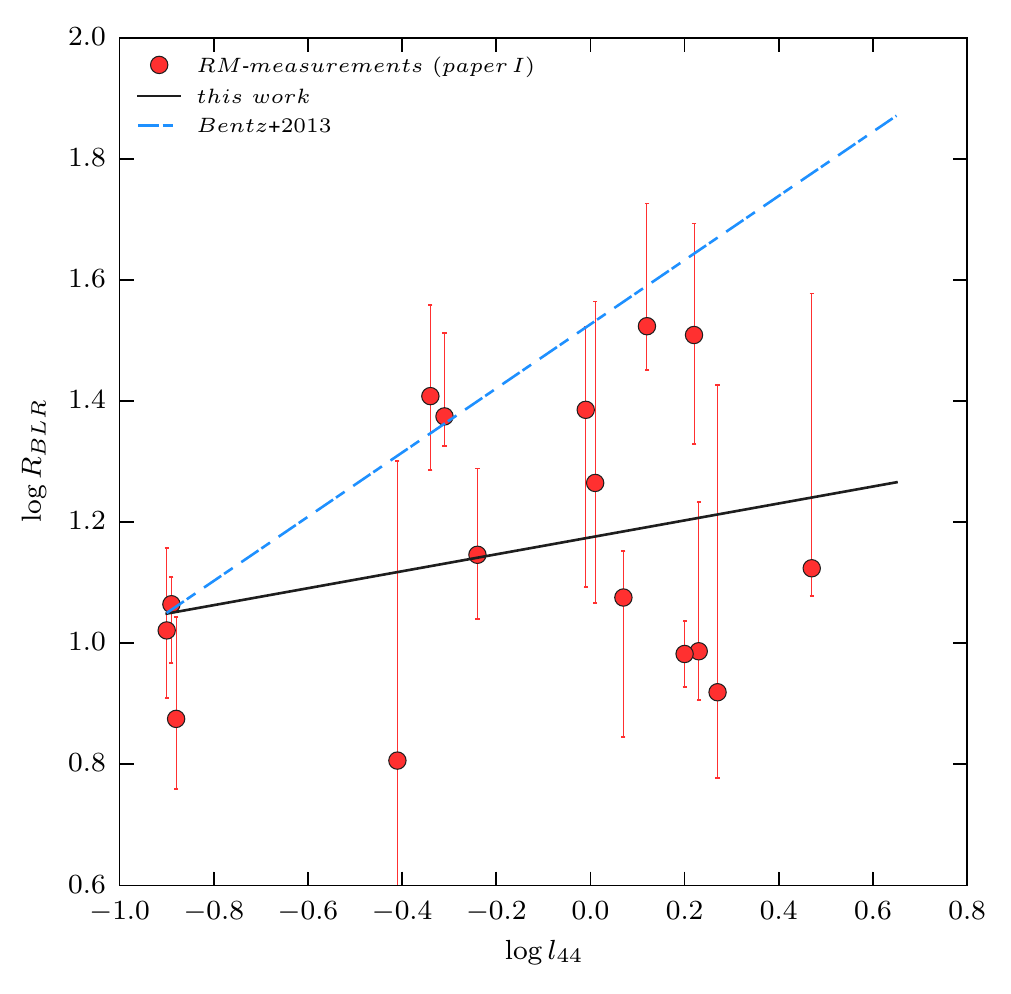}
    \caption{BLR size as a function of the optical luminosity at 5100\AA{} (in units of $10^{44}$ erg/s) for the 16
    fast accreging sources with RM-measurements that were presented in paper~{\sc I} (red points). The solid line
    represents the best-fit to this sample and corresponds to the relation assumed in this work. For comparison, we
    also show the best-fit relationship found for low Eddington ratio AGN \citep[dashed line;][]{Bentz2013}.
    }\label{fig.fitRblr}
\end{figure}

\begin{figure}
    \centering
    \includegraphics[width=0.9\linewidth]{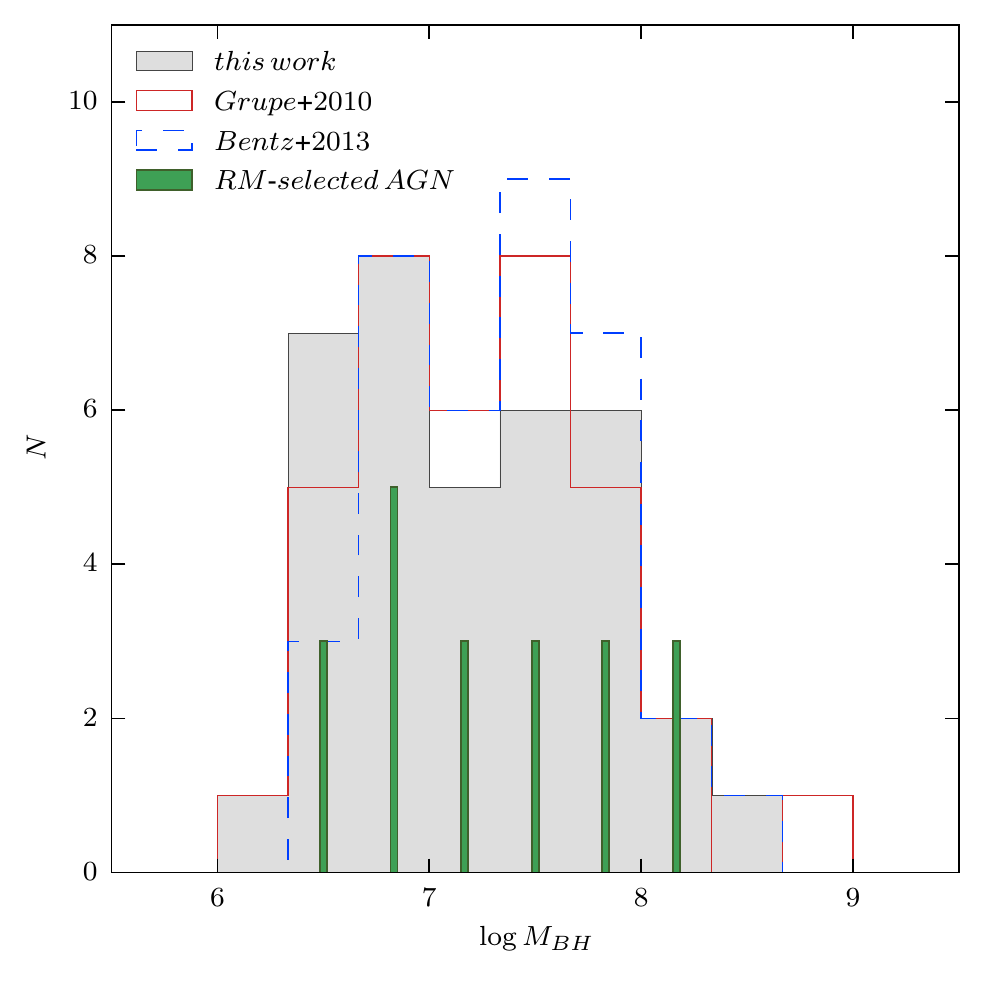}
    \includegraphics[width=0.9\linewidth]{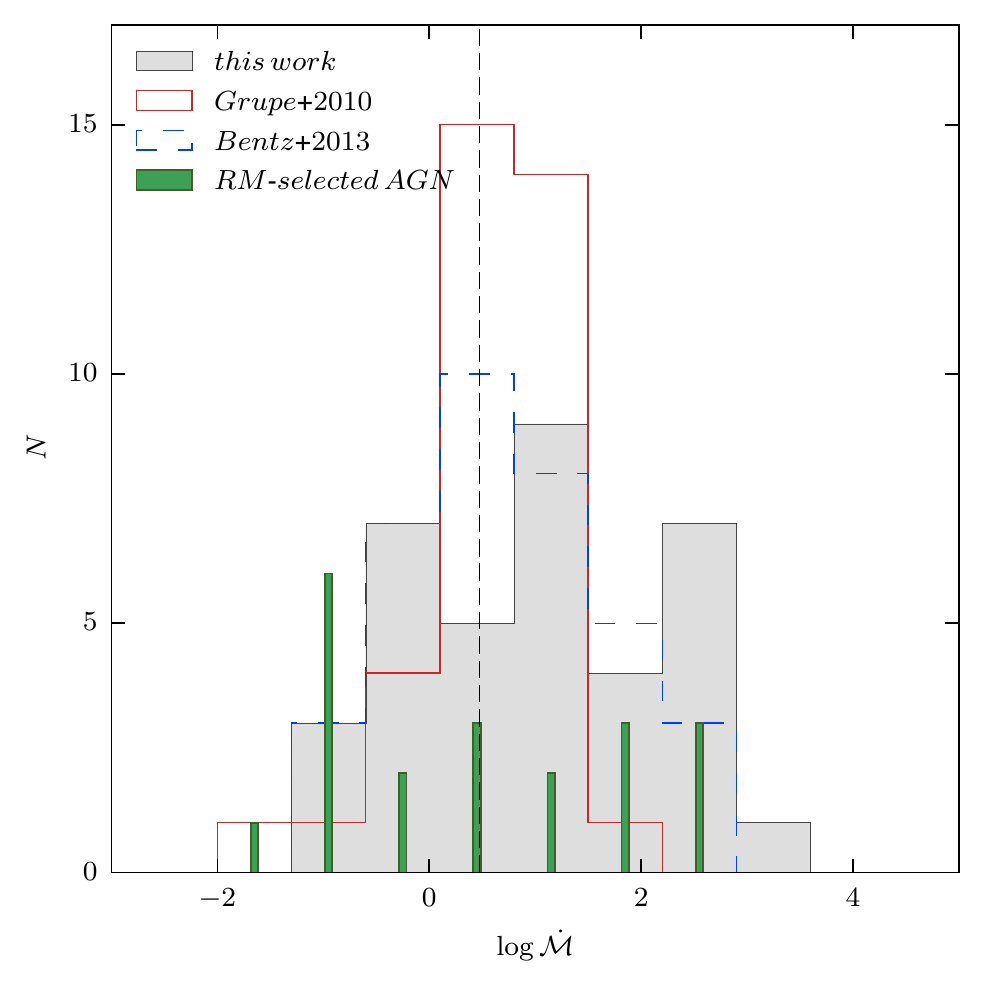}
    \caption{BH mass (top) and dimensionless accretion rate \cMdot{} (bottom)
    for the RM-selected AGN (green, thin-barred histogram) and soft-X-ray-selected
    AGN samples used in this work (grey, step-filled histogram). The solid- and dashed-line
    distributions represent the distribution of the BH mass and the accretion rate of the soft-X-ray-selected
    AGN assuming the values given in \citet{Grupe2010} and those given by the $R_{BLR}-L_{5100}$ relationship
    for low accretion rate sources, respectively \citep{Bentz2013}.
    The vertical dashed-line in the bottom panel shows the empirical Eddington ratio
    threshold at \cMdot$=3$ beyond which one needs to allow for the smaller BLR
    size for the same source luminosity.}
    \label{fig.BH_Edd_distributions}
\end{figure}

Fig.~\ref{fig.fitRblr} illustrates the use of this method by comparing the best-fit $R_{BLR}$--$l_{44}$ relationship
obtained by fitting all the fast accreting sources with RM-measurements presented in paper~{\sc I} with the best
fit relationship found for slow accreting systems \citep{Bentz2013}. The red points represent the BLR size
given by the RM measurements. It is obvious that the best-fit relationship for
slow accreting systems can not be used for sources with higher accretion rates.
In Fig.~\ref{fig.BH_Edd_distributions}, the estimated BH mass and accretion rate
obtained by our approach (gray, step-filled histogram) are compared with the most
recent $R_{BLR}$--\Lopt{} relationship
\citep{Bentz2013}, which assumes no dependency on the mass accretion rate (blue,
dashed-line histogram), and with the values used by
\citet[][red, solid-line histogram]{Grupe2010}, which were estimated using earlier
$R_{BLR}-L_{5100}$ correlations also independents of \cMdot.
The same figure also shows the distribution of the BH mass and the
dimensionless accretion rate for the RM-selected sample (green, thin-barred
histograms).

In summary, the sample consists of 55 AGN in the local Universe ($z<0.3$) of which 19 have
accurate estimations of \cMdot{} and almost 60\% (33/55) are accreting beyond
\cMdot$\geq$3. The objects in the latter group are the best
candidates to host slim ADs. While incomplete, our sample is the largest and best of its kind,
in particular in terms of a better definition of the BH mass and normalized
accretion rate, and more uniform distribution of sources across the ranges of
interest in these parameters.

\section{Data Analysis}
\label{DataAnalysis}

The IR to X-ray spectral properties of 16 out of 19 RM-selected AGN are presented in paper~{\sc I}.
For each object, we reported observations from \xmm{} that include simultaneous optical photometry and X-ray
spectrum; and IR photometry from the most recent WISE All-Sky Data Release. The intrinsic AGN
spectra were fitted with AD models in order to look for SED differences that
depend on the dimensionless accretion rate \cMdot. The fitting took into account measured BH mass and accretion rates, BH spin (only two
possibilities: $a=-1$ and $a=0.998$), host-galaxy contribution, and intrinsic reddening of the sources.
As shown in paper~{\sc I}, the primary AGN emission can be fitted by thin AD models over the range 0.2--1$\,\mu$m
regardless of the Eddington ratio. We see no evidence for extremely luminous ionizing continua, and no unusual
torus emission (albeit the large aperture WISE data), in those sources expected to be powered by slim discs.

As of early 2015, only half of the reverberation mapped slim disc candidates had X-ray
observations. Here we present, for the first time, X-ray observations of four RM-selected candidates to
have slim AD (SDSS J075101.42+291419.1, SDSS J080101.41+184840.7, SDSS J081441.91+212918.5, and SDSS J100055.71+314001.2)
that have been observed recently by XMM-Newton (PI: Shai Kaspi).
These new AGN with the highest accretion rates \cMdot$>20$ were observed for the purpose of testing the slim-disc idea in
the most extreme known cases.
We follow the work of paper~{\sc I} to fit the IR-optical-UV SED of these RM-selected AGN (see Appendix~\ref{appendix}
for a more explanations where the spectral analysis is presented). We found that the UV-optical SED of these four
RM-selected AGN can
also be fitted by thin AD models presenting moderate amount of reddening ($A_{V}\leq0.4$).
We also found that the soft X-ray emission, as well as the Comptonized power law, of these AGN with
the largest dimensionless accretion rate (\cMdot$=$20~--~200, are surprisingly similar to those from
the parent sample (\cMdot$\geq$3) that have lower accretion rates.
The spectral properties that were used in the correlation analysis for 16/19 RM-selected AGN are listed in
paper~{\sc I} and those of the new 3/19 RM-selected AGN are presented in Table~\ref{table.Xrayfit}.
As mentioned in the appendix, we were not able to carry out simultaneous optical spectrum for J080101
because of bad weather conditions and we just used the results presented in paper~{\sc I} to derived its
spectral properties.

The UV-optical-X-ray properties for the soft X-ray selected sample are taken
from \citet{Grupe2010} based on simultaneous optical and X-ray observations.
Their analysis focussed on finding differences between narrow and
broad line Seyfert 1 galaxies. For each object, they reported observations
from {\it Swift} that include simultaneous optical
(UVOT photometry) and X-ray (XRT spectrum) observations.  The optical/UV properties were determined
from a single power-law model fit to the UVOT fluxes which account for possible host-galaxy contamination and
Galactic extinction, as well as intrinsic reddening. The X-ray spectral properties were measured from the power-law
fits to the XRT data. A summary of the X-ray and optical/UV properties for the
soft-X-ray-selected AGN is given in \citet{Grupe2010}.

Following paper~{\sc I}, we inferred the torus emission by fitting the most
recent available WISE All-Sky Data Release for all the soft-X-ray-selected AGN. The torus SED was constructed from the
four WISE bands, at 3.6, 4.5, 12, and 22 $\mu$m, using only data with a signal-to-noise ratio SNR>20. The observed
infrared SED were modelled by the torus template of \citet{Mor2012} which is made of three components: a clumpy torus,
a dusty narrow line region, and a hot pure-graphite dust component which represent the innermost part of the torus. We
point the reader to paper~{\sc I} for a more detailed explanation of the fitting procedure.
Similar to paper~{\sc I}, we find no evidence that the torus SED in fast accreting AGN is different from the one in
slow accreting AGN. This means that either there is a very significant
difference in the derived tours covering factors, with fast
accreting AGN showing much smaller $L_{5\mu m}/$\Lbol{} by as much as an order of magnitude or,
that our modelling of the
UV continuum overestimates the real emission by a large factor (e.g.
saturation). We return to this issue in the following section.

\begin{table*}
  {\renewcommand{\arraystretch}{1.09}
  \caption{BH mass, intrinsic dimensionless accretion rates and the logarithm of
  the corrected monochromatic
  luminosities at 2~keV, 2500\AA{}, 5100\AA{} and 5$\mu$m in units of erg$\,$s$^{-1}$.
            The first block represents sources with RM results, while the second block corresponds to the
            soft-X-ray-selected AGN sample (see Sect.~\ref{TheSample}).}
  \label{table.resutls}
  \begin{tabular}{lccccccc}
  \toprule
  Object & $z$ & $\log{\frac{M_{BH}}{M_{\odot}}}$ & $\log{\dot{\mathcal{M}}}$ &
  $\log{\frac{L_{2\,\text{keV}}}{\text{erg/s}}}$ &
  $\log{\frac{L_{2500\mAA}}{\text{erg/s}}}$ &
  $\log{\frac{L_{5100\mAA}}{\text{erg/s}}}$ &
  $\log{\frac{L_{5\mu\text{m}}}{\text{erg/s}}}$ \\
  \midrule
  SDSS J075101.42+291419.1$^{\dagger}$ & 0.121 & 7.16 & 1.34 & 41.67 & 44.70 & 44.48 & 44.29\\
  SDSS J080101.41+184840.7 & 0.140 & 6.78 & 2.53 & 43.49 & 44.98 & 44.27 & 44.51\\
  SDSS J081441.91+212918.5 & 0.163 & 6.97 & 1.64 & 43.68 & 44.53 & 44.01 & 44.02\\
  SDSS J100055.71+314001.2 & 0.195 & 6.50 & 1.90 & 43.45 & 44.54 & 44.12 & 44.34\\
  SDSS J093922.89+370943.9 & 0.186 & 6.53 & 2.65 & 43.28 & 44.63 & 44.07 & 44.47\\
  Fairall 9 & 0.047 & 8.09 & -0.70 & 43.63 & 44.60 & 43.98 & 44.57\\
  IRAS F12397+3333 & 0.044 & 6.79 & 2.94 & 43.23 & 44.51 & 44.23 & 43.82\\
  PG 0844+349 & 0.064 & 7.66 & 0.82 & 43.52 & 44.69 & 44.22 & 44.17\\
  PG 2130+099 & 0.063 & 7.05 & 1.75 & 43.23 & 44.75 & 44.20 & 44.62\\
  PG 1229+204 & 0.063 & 8.03 & -1.04 & 43.43 & 44.23 & 43.70 & 44.09\\
  Mrk 817 & 0.031 & 7.99 & -0.88 & 43.19 & 44.12 & 43.74 & 44.14\\
  Mrk 279 & 0.030 & 7.97 & -0.86 & 43.40 & 44.15 & 43.71 & 43.69\\
  Mrk 290 & 0.029 & 7.55 & -0.82 & 42.85 & 43.89 & 43.17 & 43.58\\
  Mrk 79 & 0.022 & 7.84 & -0.60 & 42.39 & 43.79 & 43.68 & 43.86\\
  Mrk 1511 & 0.034 & 7.29 & -0.35 & 42.84 & 43.64 & 43.16 & 43.48\\
  Mrk 590 & 0.026 & 7.55 & -0.28 & 42.65 & 43.59 & 43.50 & 43.39\\
  Mrk 110 & 0.035 & 7.10 & 0.85 & 43.63 & 44.35 & 43.66 & 43.76\\
  Mrk 382 & 0.033 & 6.50 & 1.20 & 42.84 & 43.65 & 43.12 & 43.22\\
  Mrk 335 & 0.026 & 6.93 & 1.39 & 43.31 & 44.29 & 43.76 & 43.97\\
  \midrule
  RX J2301.6-5913 & 0.149 & 8.48 & -1.07 & 44.07 & 44.36 & 44.27 & 44.67\\
  RX J1413.6+7029 & 0.107 & 7.80 & -0.27 & 43.41 & 43.54 & 43.89 & 44.03\\
  RX J0311.3-2046 & 0.070 & 7.85 & -0.20 & 43.41 & 44.19 & 44.00 & 44.18\\
  RX J0437.4-4711 & 0.052 & 7.75 & -0.06 & 43.20 & 44.04 & 43.97 & 43.87\\
  RX J0100.4-5113 & 0.062 & 7.57 & 0.34 & 43.08 & 44.10 & 44.00 & 44.19\\
  RX J0319.8-2627 & 0.076 & 7.60 & 0.43 & 43.02 & 44.21 & 44.09 & 43.94\\
  RX J2146.6-3051 & 0.075 & 7.54 & 0.44 & 43.15 & 44.17 & 44.02 & 44.00\\
  RX J0105.6-1416 & 0.070 & 7.18 & 1.26 & 43.51 & 44.16 & 44.09 & 44.14\\
  RX J0859.0+4846 & 0.083 & 7.34 & 1.30 & 43.48 & 44.62 & 44.32 & 44.26\\
  RX J0128.1-1848 & 0.046 & 7.18 & 1.15 & 43.37 & 44.04 & 44.01 & 43.90\\
  RX J1007.1+2203 & 0.083 & 6.61 & 1.64 & 42.53 & 43.65 & 43.58 & 43.68\\
  RX J2258.7-2609 & 0.076 & 6.99 & 1.91 & 43.13 & 44.30 & 44.27 & 43.89\\
  RX J2242.6-3845 & 0.221 & 6.96 & 2.27 & 43.67 & 44.61 & 44.47 & 44.51\\
  RX J2217.9-5941 & 0.160 & 6.67 & 2.40 & 42.84 & 44.32 & 44.17 & 44.68\\
  RX J1355.2+5612 & 0.122 & 6.44 & 2.82 & 43.18 & 44.30 & 44.14 & 44.16\\
  RX J2317.8-4422 & 0.132 & 6.36 & 2.85 & 42.96 & 44.16 & 44.05 & 44.06\\
  RX J1304.2+0205 & 0.229 & 6.63 & 2.88 & 43.39 & 44.58 & 44.43 & 44.72\\
  RX J1319.9+5235 & 0.092 & 6.30 & 2.97 & 42.81 & 44.24 & 44.06 & 43.68\\
  Mrk 493 & 0.032 & 6.11 & 2.84 & 42.25 & 43.79 & 43.71 & 43.63\\
  Mrk 684 & 0.046 & 6.54 & 2.41 & 42.94 & 44.11 & 44.00 & 43.82\\
  Mrk 841 & 0.036 & 8.00 & -0.86 & 43.07 & 43.95 & 43.77 & 43.93\\
  Mrk 1048 & 0.042 & 8.11 & -0.63 & 43.44 & 44.30 & 44.07 & 44.20\\
  Mrk 141 & 0.042 & 7.53 & 0.01 & 42.77 & 43.49 & 43.72 & 43.69\\
  Mrk 142 & 0.045 & 6.70 & 1.48 & 42.84 & 43.72 & 43.60 & 43.62\\
  AM 2354-304 & 0.033 & 7.05 & 0.80 & 42.70 & 43.56 & 43.60 & 43.41\\
  MS 2254-36 & 0.039 & 6.66 & 1.65 & 42.82 & 43.81 & 43.66 & 43.80\\
  NGC 4593 & 0.009 & 7.81 & -0.52 & 41.51 & 43.44 & 43.74 & 43.11\\
  Fairall 1119 & 0.055 & 7.70 & -0.41 & 43.15 & 43.31 & 43.66 & 43.73\\
  Fairall 1116 & 0.059 & 7.88 & -0.15 & 43.25 & 44.26 & 44.08 & 44.37\\
  Fairall 303 & 0.040 & 6.58 & 1.45 & 42.79 & 43.62 & 43.42 & 43.40\\
  Ton 730 & 0.087 & 7.61 & 0.20 & 43.40 & 44.36 & 43.96 & 44.05\\
  CBS 126 & 0.079 & 7.30 & 1.02 & 43.07 & 44.35 & 44.09 & 44.03\\
  ESO 301-G13 & 0.059 & 7.08 & 1.10 & 43.18 & 43.96 & 43.85 & 44.08\\
  EXO 1627+40 & 0.272 & 6.70 & 2.52 & 44.11 & 44.50 & 44.29 & 44.54\\
  KUG 1618+410 & 0.038 & 6.80 & 1.19 & 42.21 & 43.64 & 43.53 & 43.05\\
  VCV 0331-37 & 0.064 & 6.85 & 1.41 & 43.06 & 43.94 & 43.75 & 43.34\\
\bottomrule
\multicolumn{8}{l}{$^{\dagger}$The short X-ray exposure time does not allow to obtain good quality spectra above 3 keV.}\\
\end{tabular}
}
\end{table*}

\section{Results and Discussion}
\label{Results}

\subsection{Correlation Analysis}
All spectral properties used in this paper are taken either from the spectral analysis presented in the works of
\citet{Grupe2010}, paper~{\sc I} or the analysis presented in Appendix~\ref{appendix}.
We then explore the connection between the dimensionless accretion rate \cMdot{}, the X-ray and the infrared
loudness\footnote{The X-ray loudness is defined as the optical-to-X-ray spectral slope
$\alpha_{OX}=-0.3838\log{\left[
\frac{L_{\nu}(2\text{keV})}{L_{\nu}(2500\text{\AA{}})}\right]}$ \citep{Tananbaum1979}}
($\alpha_{OX}$ and $\alpha_{IO}$, respectively) in low-to-intermediate luminosity AGN, highlighting the slim AD candidates.

There are several fundamental differences between previous correlation analysis \citep[][and references therein]{Vasudevan2009,Grupe2010} and
this work. First, the BH mass estimated is based on a more realistic $R_{BLR}$--\Lopt{} relationship which leads to more accurate
accretion rates. Second, the large number of sources per Eddington ratio interval, reduce the uncertainty due to variability of such
sources especially at X-ray energies. Third, the correlation analysis is based on
a Bayesian approach instead of a simple linear correlation analysis.

We adopted a new methodology to analyse the connection between the dimensionless mass accretion rate \cMdot{}
and the optical-to-X-ray spectral slope \alphaox{} (see Section~\ref{XrayLoudness}), as well as its connection with the
IR-to-optical spectral slope \alphaio{} (see Section~\ref{IRLoudness}).
The paired parameters (e.g. \alphaox{}--\cMdot{}) that represent a
collection of individual measurements $\{(x_i,y_i)\}$ (e.g. $\{(\alpha_{{OX}_{i}},\dot{\mathcal{M}}_{i})\}$)
is characterized by a two-step Bayesian approach. We use a Markov Chain Monte Carlo (MCMC)-based algorithm
to study the ``level'' of correlation between both parameters and to characterize it.

First, we evaluate the degree of correlation between the two parameters by inferring
the posterior probability density function (PDF) of the correlation coefficient assuming the following priors.
The individual measurements $(x_i,y_i)$ are independent with Gaussian distributed uncertainties and the distribution of
the paired parameters is given by a bivariate normal distribution
\begin{equation*}
\begin{array}{crl}
x_i & \mathcal{P}(x_i) & =\mathcal{N}(\mu_{x_i}=x_i,\sigma_{x_i}=1/\delta_{x_i}^2)\\
y_i & \mathcal{P}(y_i) & =\mathcal{N}(\mu_{y_i}=y_i,\sigma_{y_i}=1/\delta_{y_i}^2)\\
(x_i,y_i)& \mathcal{P}(x_i,y_i  ) & =\mathcal{N}(\mu(\mu_{x_i},\mu_{y_i}),\Sigma(\sigma_{x_i},\sigma_{y_i},\rho))\\
\end{array}
\end{equation*}
where $\delta_{x_i}$ and $\delta_{y_i}$ are the measured uncertainties on $x_i$ and $y_i$, respectively; and
$\rho$ is the correlation coefficient with a uniform prior distribution defined between -1 (anti-correlated with
no scatter) and +1 (correlated with no scatter).

For those cases that show signs of correlation, we characterize
the parameters that govern such correlation through modelling the
collection of individual measurements $\{(x_i,y_i)\}$ with three different functional forms:
a line
\begin{equation}
f(x)=\beta x + \alpha_0
\end{equation}
a broken power-law
\begin{equation}
\label{eq.bknpl}
f(x)=\left
    \{\begin{array}{cc}
        \alpha_0 & x<\dot{\mathcal{M}}_c \\
        \alpha_0 + \beta \log{\frac{x}{\dot{\mathcal{M}}_c}} & x\geq \dot{\mathcal{M}}_c \\
    \end{array}
    \right.
\end{equation}
and a sigmoid given by
\begin{equation}
\label{eq.sigmoid}
f(x) = \alpha_0 + \frac{\Delta \alpha_0}{1+e^{-\beta \log{\frac{x}{\dot{\mathcal{M}}_c}}}}
\end{equation}
Assuming also independence between the individual measurements and Gaussian distribution for
their uncertainties, we infer the posterior PDF of each model parameter: $\{\alpha_0,\beta\}$,
$\{\alpha_0,\beta, \dot{\mathcal{M}}_c\}$ or $\{\alpha_0,\Delta \alpha_0, \dot{\mathcal{M}}_c, \beta \}$ (for the line, the broken power-law and the sigmoid
function, respectively) assuming that
\begin{equation*}
y_i = f(x_i) + \mathcal{N}(\mu=y_i, \sigma=1/\delta_{y_i}^2).
\end{equation*}

After testing for several possible priors (Gaussian, Exponential and Uniform probabilities), we found that the
posterior PDF of the model's parameter is insensitive to its prior distribution.
Finally, once the posterior PDFs are inferred (i.e. the model is fitted), we evaluate the goodness of the fit
through the statistic $p$ which uses the Freeman-Turkey criterion. Using the model's parameter posterior PDFs, we
generate a family of $N$ ``expected values'' (and hence a family of $N$ simulated values), to infer the
probability distribution of the difference between both the observed and simulated to the expected values.
On average we expected the difference between them to be zero;
hence, $p$ is simply the fraction of simulated discrepancies that are larger than their corresponding observed
discrepancies. Therefore, if $p$ is very large ($p>0.975$) or very small ($p<0.025$) the model is not
consistent with the data.

Throughout this work, the listed value for the correlation coefficient $\rho$, the model's parameter
and the goodness of the fit $p$ correspond to the most likely value according to its posterior PDF, and the
errors represents the 2.5$^{\text{th}}$ and 97.5$^{\text{th}}$ percentiles of that probability.

\subsection{X-ray loudness, \alphaox}
\label{XrayLoudness}
We applied the method, presented in the previous section, to analyse the connection between the X-ray loudness,
\alphaox{}, and \cMdot{}.

Some correlations between \alphaox{} and other spectral properties have been
reported. While, it is widely accepted that \alphaox{} correlates with the X-ray
bolometric correction factor and the hard X-ray photon index
\citep[][]{Vasudevan2009b,Marchese2012,Jin2012},
a possible correlation between \alphaox{} with the Eddington ratio is still under debate.
Some earlier studies \citep{Vasudevan2009b,Fanali2013,Plotkin2016} report a lack of a correlation
between \alphaox{} and \mdot{}, but \citet{Grupe2011} claim that they are strongly correlated.
Most important, perhaps, is the ability to separate those correlations linking
\alphaox{} to source luminosity, over a large luminosity range
\citep[e.g.][]{Vasudevan2009,Grupe2011} from those comparing \alphaox{} with the
normalized accretion rate.
To understand the nature of slim AD systems, we will focus solely on the
correlation with \cMdot{}. Our main concern is to explore whether the fast
accreting systems follow the same connection between \alphaox{} and
\cMdot{} over a limited range of luminosities.
We note, again, that in our sample we can only constrain \Lopt{}, not Lbol{}.

\begin{figure}
    \centering
    \includegraphics[width=0.9\linewidth]{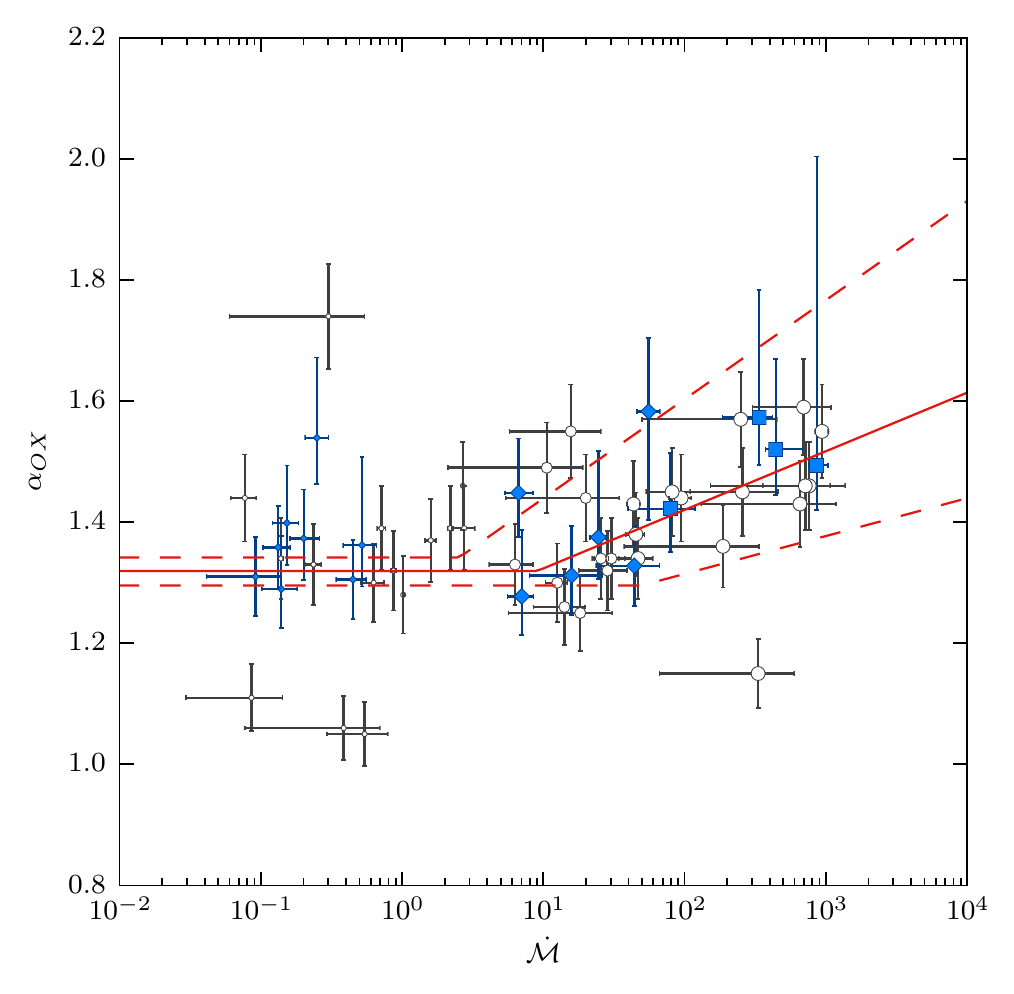}\\
    \includegraphics[width=0.9\linewidth]{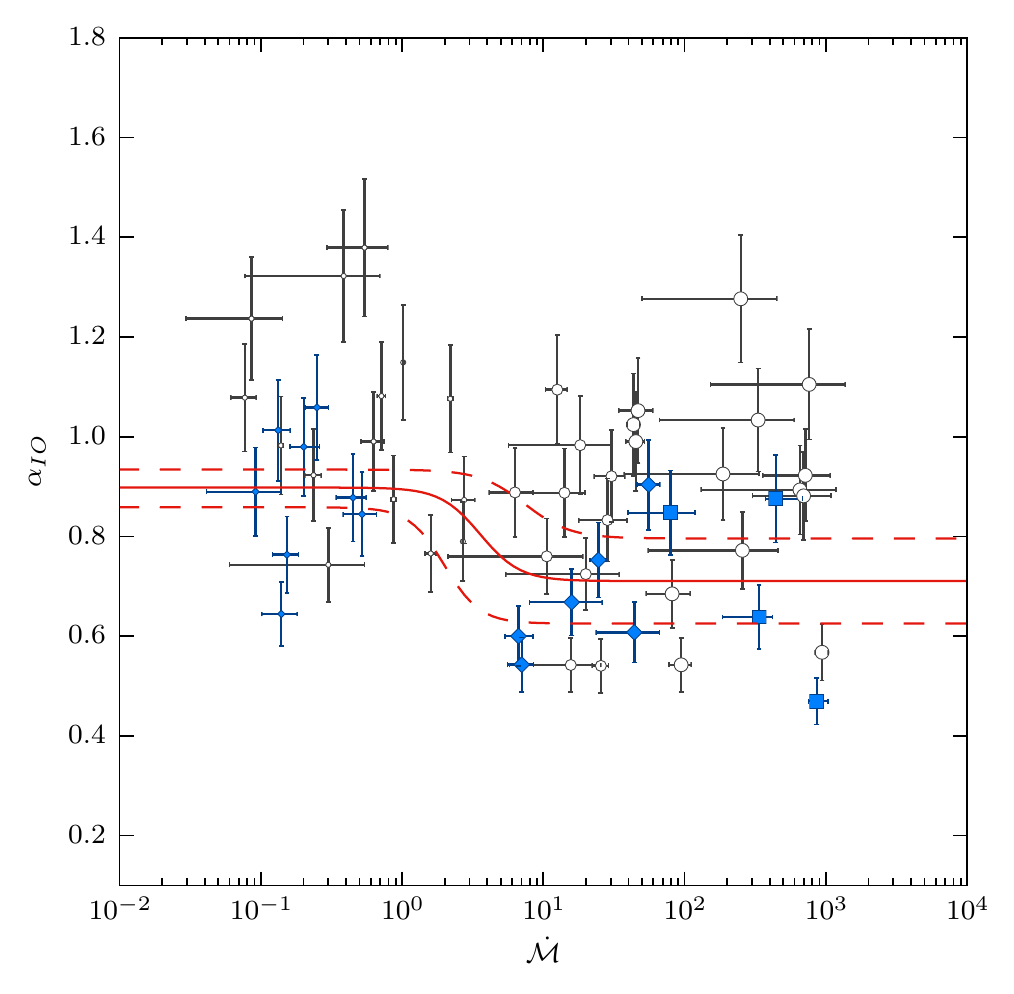}
    \caption{X-ray (\alphaox) and IR (\alphaio) loudness as a function of the Eddington ratio for both RM-selected
    AGN sample (blue, filled diamonds) and soft-X-ray-selected AGN sample (grey, empty circles). The symbol size scales
    with the mass accretion rate bins: \cMdot$<$3.0, $3.0\leq$\cMdot$<$\cMdotcrit, and \cMdot$\geq$\cMdotcrit. The
    diagrams show our best fit model for the two correlations, a broken power law at the top and a sigmoid function
    at the bottom.}
    \label{fig.alphaOX}
\end{figure}

After re-estimating the BH mass, and hence the Eddington ratio, by assuming the
new dependence of $R_{BLR}$ on \Lopt{} (section 2.1), and adding accurate measurements of
the accretion rate (RM-selected AGN), we find that \alphaox{} is \emph{weakly
correlated} with \cMdot{} ($\rho=0.31^{+0.20}_{-0.25}$). As we can see in the Fig.~\ref{fig.alphaOX},
the high Eddington ratio tail, i.e. \cMdot$\gtrsim10$, displays the larger
spread in \alphaox{} and appears to be slightly shifted to
higher values.
When the sample is limited to those AGN that are accreting below \cMdot$< 10$ (28/59) the
analysis shows a lack of correlation ($\rho=-0.04^{+0.37}_{-0.36}$). This suggests that
the \emph{weak} $\alpha_{OX}$-\cMdot{} \emph{correlation} is dominated by the fast accreting AGN. Indeed,
when the sample is limited to those AGN that are accreting beyond
\cMdot$\geq10$  (31/59), we find that \alphaox{}
is strongly correlated with \cMdot{} with $\rho=0.42^{+0.10}_{-0.11}$.
This results confirms what it is usually found in the literarue, i.e. the X-ray
loudness does not correlate with the dimensionless accretion rate. This result is at odds with that of \citet{Grupe2011} who find a strong correlation between the X-ray loudness and
the Eddington ratio. However, the Eddington ratio derived by \citet{Grupe2011} are based on the bolometric luminosity as given
by the best-fit power-law model to simultaneous X-ray spectra and UVOT photometry which introduces a non-systematic error on the
inferred Eddington ratio.

\begin{figure}
    \centering
    \includegraphics[width=\linewidth]{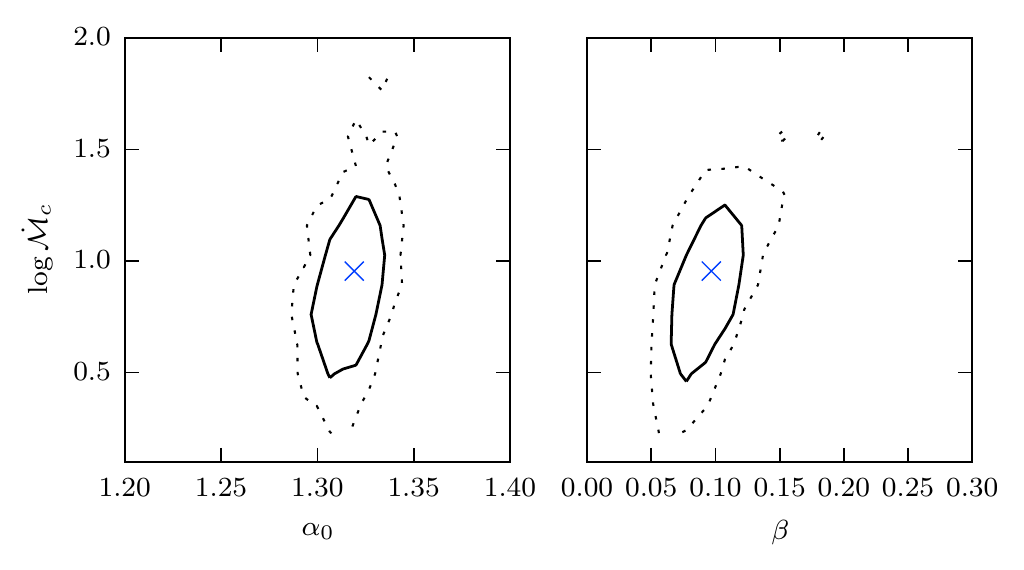}
    \caption{The 68 percent and 95 percent confidence intervals from the MCMC
    analysis of \alphaox{}--\cMdot{}. Contours show the constraints on the transition accretion rate with both
    the most likely \alphaox{} for low accreting AGN (\emph{left}) and the right side shows the best $\beta$ in the
    broken power-law model. The cross point to the most likely value of each model's parameter. }
    \label{fig.contour}
\end{figure}

Our approach to this dependence is different and assumes that \alphaox{} is different in low
and high accretion rate AGN and that there is a smooth transition between these value which may be related
to some new physics (e.g. the onset of slim disc accretion). We modelled the dependence of \alphaox{} on the
accretion rate \cMdot{} by fitting a broken power law (Eq.~\ref{eq.bknpl}), as well as
a sigmoid function (Eq.~\ref{eq.sigmoid}),
to the data. Both models give satisfactory fit to the data but the broken power law is slightly better and hence
is used for the rest of the analysis. Fig.~\ref{fig.alphaOX} shows our determination for the transition
accretion rate (\cMdot$_{c}$ in Eq.~\ref{eq.bknpl}),
the most likely \alphaox{} for low accreting sources ($\alpha_0$ in
Eq.~\ref{eq.bknpl}) and the transition between low and high \alphaox{} regime
which is regulated by the parameter $\beta$ in Eq.~\ref{eq.bknpl}.
As shown in Fig.~\ref{fig.contour}, the model's parameters (\cMdot$_{c}$, $\alpha_0$ and $\beta$)
are well constrain.
The posterior probability density gives a transition accretion rate of
$\dot{\mathcal{M}}_{c}=8.9^{+40.0}_{-6.4}$ and a X-ray
loudness of $\alpha_{OX} = \alpha_0 = 1.32^{+0.02}_{-0.03}$ for objects that are
accreting below \cMdot$<$\cMdot$_{c}$
which is significantly lower than the most likely value for the highest accretion rates
\alphaox$=1.49^{+0.05}_{-0.05}$.
We also tried to fit the data with a line, with no transition point, but that did not result in a
satisfactory fit, $p=0.015$.

While the uncertainty on the preferred value of the transition accretion rate is
large, because of the small number of objects and the large uncertainties on the
accretion rate, it is clear that the value obtained from fitting our sample is
considerably lower than \cMdot$_{c}=$50 preferred by \citet{Mineshige2000}, but
consistent with \cMdot$_{c}=$20 suggested by \citet{Watarai2001}.
The transition accretion rate found here is also consistent with
those found by \citet{DuPu2015} and \citet{DuPu2016}, which are also based on
observational constrains (\cMdot$_{c}=13.8^{+19.6}_{-8.1}$ and
\cMdot$_{c}=11.19^{+2.29}_{-6.22}$, respectively).
The minimum accretion rate of \cMdotcrit{}$=$3 used by us to define slim AD
candidates is within the error bar, but at the lower end of the confidence level.
As pointed out in \citet{DuPu2015}, there must
be a smooth transition from thin to slim discs and, wherefore, relatively large range in
\cMdot{} from flat to linear dependence on \cMdot.

\subsection{Infrared loudness, \alphaio}
\label{IRLoudness}

The re-emitted IR radiation can also be used as a proxy of the primary nuclear emission.
In this section we use the IR (torus) emission to carry an analysis of the optical-IR relation,
similar to the UV-X-ray relation discussed in the previous section.
Figure~\ref{fig.alphaOX} shows the dependence on the dimensionless accretion
rate of the IR loudness which is defined as the infrared-to-optical spectral slope
\begin{equation}
    \alpha_{IO} = -0.7687 \log{\frac{L_{\nu}(2500\text{\AA{}})}{L_{\nu}(5\mu\text{m})}}.
\end{equation}
In contrast to the X-ray loudness, we find that there is no correlation between \alphaio{} and
\cMdot{} when the sample is split into lower (\cMdot$<10$) and higher (\cMdot$\geq10$) mass
accreting AGN ($\rho\sim 0$). However, when the whole sample is considered we find a weak
correlation ($\rho=-0.25^{+0.22}_{-0.09}$). As we can see in Figure~\ref{fig.alphaOX},
those sources accreting beyond \cMdotcrit{} appear to be shifted to lower \alphaio{} values.
We modelled the correlation by fitting the data to a sigmoid function
(Eq.~\ref{eq.sigmoid}). This enables us to determine the transition accretion
rate (\cMdot{}$_{c}$ in Eq.~\ref{eq.sigmoid}), as well as the most
likely \alphaio{} for low ($\alpha_0$ in Eq.~\ref{eq.sigmoid}) and high
($\alpha_0+\Delta\alpha_0$ in Eq.~\ref{eq.sigmoid})
accretion rate AGN.
We found that the transition accretion rate is consistent with that given by the
\alphaox{}--\cMdot{} correlation (\cMdot$_{c}=3.6^{+5.0}_{-1.2}$). The posterior
probability density gives an IR loudness of \alphaio{}$=\alpha_0=0.91^{+0.03}_{-0.04}$
for objects that are accreting below \cMdot$<$\cMdot$_{c}$ and a shift between low
and high accreting AGN of $\Delta\alpha_0=-0.19^{+0.04}_{-0.05}$.
This means that for a given UV monochromatic luminosity, the torus luminosity appear to be slightly
lower for high accreting AGN (see Figure~\ref{fig.medianSED}).

\subsection{Median SED}
\begin{figure}
\includegraphics[width=\linewidth]{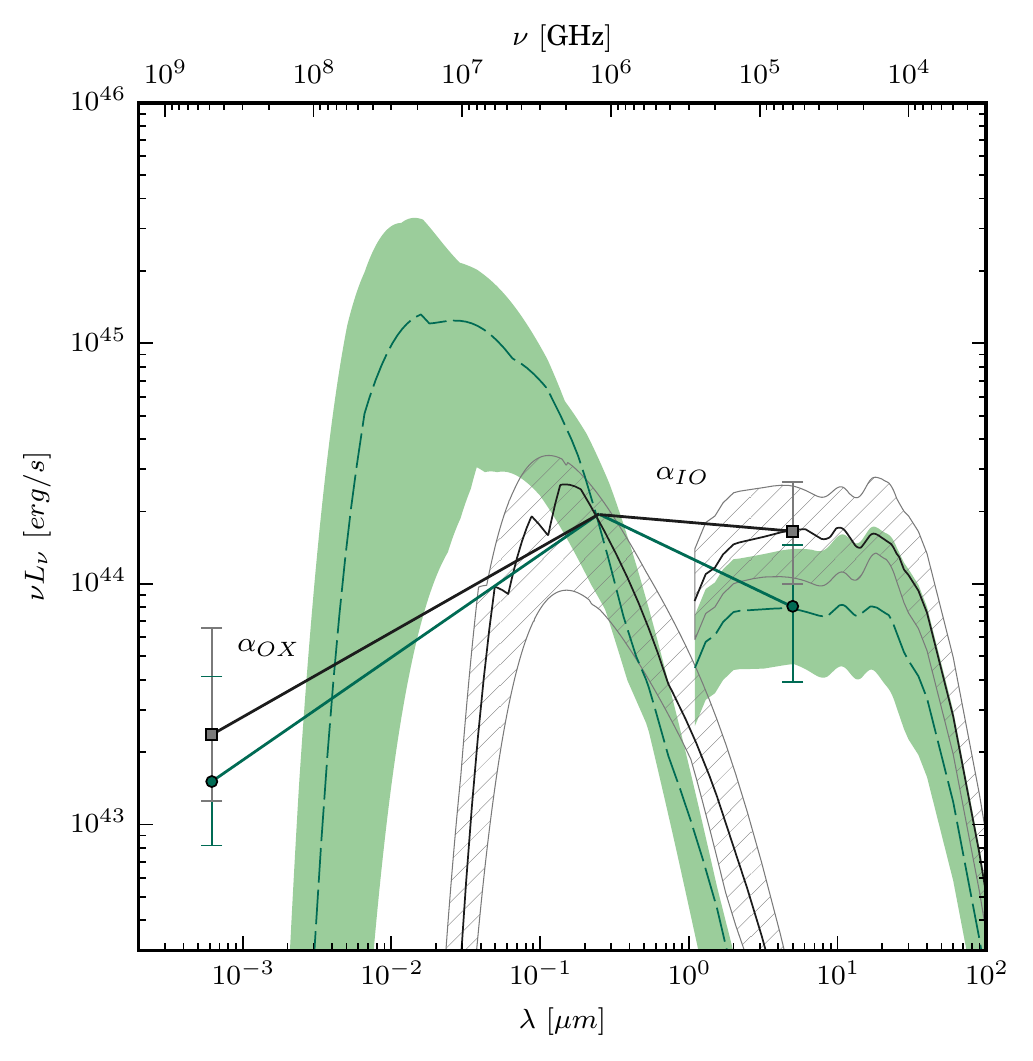}
\caption{Median AGN SEDs normalized at 2500\AA{} for the two groups: green, dashed line represents
   sources accreting at \cMdot$\geq$\cMdotcrit, and black, solid line sources accreting at \cMdot$<$\cMdotcrit.
   The dashed areas show the 25$^{th}$ and 75$^{th}$ percentiles.
   The square (\cMdot$<$\cMdotcrit{}) and circle (\cMdot$\geq$\cMdotcrit{}) points represent
   the median luminosity at 2keV and 5$\,\mu$m. For each Eddington ratio bin, there are two
   \emph{horitzontal} lines that represents \alphaox{} and \alphaio{}.
   }
\label{fig.medianSED}
\end{figure}

Fig.~\ref{fig.medianSED} shows the median primary AGN and torus SED normalized at 2500\AA{} for two
ranges: \cMdot$<$\cMdotcrit{} and \cMdot$\geq$\cMdotcrit{}
(\cMdotcrit$=$3; solid and dashed lines, respectively). The median primary AGN
SED was constructed from the best-fitted thin
AD models to the RM-selected sample. To construct the median torus SED we used the best-fitted
torus templates of the entire sample (i.e. RM- and soft-X-ray-selected samples). The four straight
lines represent the median optical-to-X-ray and IR-to-optical spectral
slopes (note that the diagram shows $\nu L_{\nu}$ and the spectral slopes -\alphaox{}, \alphaio{}-
are defined through monochromatic luminosities).
In general, the torus and the disc corona in the fastest accreting systems, i.e.
\cMdot$\gtrsim$10, are less efficient in reprocessing the primary AGN radiation
having (i.e. higher \alphaox{} and lower \alphaio{}).

In paper {\sc I}, we found a relatively low ratio of total AGN luminosity,
produced mostly in the UV part of the disc SED, and the reprocessed radiation by
the torus for fast accreting systems, $L_{torus}/L_{AGN}\sim 10^{-1}-10^{-2}$.
This ratio is not too different from the one predicted by several theoretical
estimates \citep{Kawakatu2011} which
take into account both the saturated luminosity of the slim disc and the strong
angular dependence of the emitted radiation. However, the surprising similarity
of \Lopt{}$/L_{torus}$ in AGN with slow and fast accretion rates (thin and slim
ADs) suggests either a fine tuning of the parameters or some wrong assumptions
about the physics of slim ADs.
The IR loudness measured here, which shows how the torus of fast accreting systems
are less efficient in reprocessing the emitted intrinsic AGN radiation, is a
manifestation of this issue.

A simple way to appreciate this difficulty is to note that the bolometric correction
factor at \Lopt{} is very different for thin and slim ADs even when saturation is taken
into account. This is illustrated in Fig.~6 of paper~{\sc I} where we show this
correction factor to be about 10 for thin ADs and 100 or more for slim discs.
Thus a certain \Lopt{} will indicate a slim disc luminosity which is at least an
order of magnitude larger than the one for a thin disc. However, the uniform
\alphaio{} we measure suggest that $L_{torus}/$\Lopt{} is basically identical
for both groups. This would mean that either the torus geometrical covering factor in
systems powered by slim ADs is 10 times or more smaller, or that the angular
dependence of the emitted SED is such that, compared with thin disc systems, the
torus abserved only 10\% of the emitted radiation.
At first, this by itself seemed to be unlikely because it requires such fine
tuning. However the inclusion of self-shadowing effects and/or highly energetic
winds in slim AD, which must dramatically affect the re-emitted IR emission,
might help to make such fine tuning possible.


\section{Conclusions}
\label{Conclusions}

We conducted an investigation of various spectral properties in two AGN samples
covering a limited luminosity range. The RM measurements and the new estimates
of $R_{BLR}$ introduced here, allow the most accurate BH mass estimates, and hence Eddington ratios.
We implemented a Bayesian approach which enables us to infer the posterior probability density of the correlation
coefficient ($\rho$) used to evaluate the degree of correlation between two spectral properties, \alphaox{} and \alphaio{},
and the dimensionless accretion rate of the sources, \cMdot{}.
The main results of our study can be summarized as follows:

\begin{enumerate}
\item The optical-to-UV SED of all sources, including those with the highest mass accretion rates (\cMdot$>$20),
    can be fitted  with thin AD models, as claimed in our previous work (paper~{\sc I}).
\item We do not find any correlation between the X-ray loudness (\alphaox) and
    the dimensionless accretion rate (\cMdot), up to a very large value of \cMdotcrit$\sim 10$.
    However, the highest Eddington ratio (\cMdot$\geq$\cMdotcrit) sources appear
    to show systematically larger values of \alphaox{}. The correlation claimed in earlier works
    which extends to much lower Eddington ratios is due to a small number of
    very large \cMdot{} sources that were compared with low \cMdot{} sources, as
    well as to an inaccurate  estimation of the mass accretion rate.
\item We defined a new IR-to-optical spectral index, \alphaio{}, which is a
    measure of the reprocessing efficiency of the central torus.
    We found that the distribution of \alphaio{} of the highest Eddington ratio group appear to
    be shifted to lower values. This finding seems to contradict several of the
    assumptions used to obtain the good fit to the observed SEDs in most of the
    sources. We suggest that additional physical processes that act to reduce the
    extreme UV radiation are at work in fast accreting AGN related to photon trapping, strong winds,
    and perhaps other yet unknown physics processes. The alternative
    explanation, which we consider unlikely, is extremely small covering factor for high
    \cMdot{} sources \citep[see][]{Castello2016}.
\end{enumerate}

Our new results, failing to recover the predicted differences between the SEDs
of slim and thin ADs, within the extensive
wavelength range considered here, hint to a sever problem in present slim AD
theoretical models. Additional physics, occurring in the nuclear regions of high
accreting AGN and responsible for making the SED of slim ADs consistent with that of
thin discs, must be at work. One such process could be powerful winds as slim AD simulations suggests.
In this scenario winds are responsible for carrying away a significant amount of
the energy from the inner-most region of the AD,
decreasing its radiation in the UV part of the spectrum, in particular the UV ionizing
radiation responsible for high excitation
emission lines, as well as reduce the Comptonized emission or perhaps quenching the disc corona.
This will allow reconciliation
of our observations (IR/optical/UV/X-ray) with slim disc theory and simulations. If such winds originating in the innermost
part of the slim disc they can affect the dusty torus structure in a way which
is different from what is known in slow accreging systems.
Such an effect has never been investigated.

\section*{Acknowledgements}

We would like to thank the referee, Chris Done, for helpful
comments that greatly improved the clarity of the paper.
Funding for this work has been provided by a joint ISF-NSFC grant number 83/13.
HN acknowledges the support of ISSI during a work meeting in 2015.
LCH acknowledges support from the Chinese Academy of Science (grant No.
XDB09030102), National Natural Science Foundation of China (grant No. 11473002),
and Ministry of Science and Technology of China (grant No. 2016YFA0400702).
Bian W.H. thanks the support from the National Science Foundations of China
(No. 11373024).

%



\bibliographystyle{mnras}
\bibliography{references} 

\appendix

\section{New simultaneous broadband SED}
\label{appendix}
The following is a detailed spectral analysis of four SEAMBHs with the highest Eddington ratio up to now
(SDSS J075101.42+291419.1, SDSS J080101.41+184840.7, SDSS J081441.91+212918.5, and SDSS J100055.71+314001.2) that
have been recently observed with \xmm{}. The structure of this appendix is as follows:
the new X-ray observations are described in Sect.~\ref{XrayData}; the optical-to-UV SED is presented in
Sect.~\ref{fittingThinADmodels}; finally, the X-ray spectral analysis, as well as the principal conculsions
are discussed in Sect.~\ref{XrayAnalysis}.

\subsection{New XMM-Newton Observations}
\label{XrayData}

The sources J075101, J080101, J081441 and J100055 were observed with \xmm{} between May and October of 2015.
During all the observations, the European Photon Imaging Cameras (EPIC) pn and MOS were operated in full window mode.
The observation data files (ODFs) were processed following the standard procedures to produce calibrated event lists
using the Science Analysis System (SAS 14.0.0). We defined the source extraction region to be circular
with radii ranging from 30\arcsec{} to 35\arcsec{} centred on the bright pixel closer to
the optical source position. We used single- and double-pixel events for all observations.
The background was selected from nearby circular regions free of sources and
preferentially on the same CCD chip. Spectral response files were generated using
the SAS tasks {\sc rmfgen} and {\sc arfgen}. The {\sc epatplot}
SAS task was used to test for the presence of pile up. The EPIC X-ray spectra of all
the observations were found to be free from the effects of pile-up.

The sources were also observed with the Optical Monitor (OM) abroad of \xmm{},
providing us with optical to UV photometry simultaneously to the EPIC observations. These
observations were reduced using the {\sc omichain} pipeline.
Point source and extended source identification is automatically performed as part of
this pipeline, and the point source corresponding to the nucleus was selected using
the images generated for each waveband, ensuring that the identified optical or UV
counterpart was coincident with the source region used for the X-ray spectrum. The
optical/UV photometry results were extracted from the {\sc swsrli} results files. To prepare
the OM data, the \emph{om\_filter\_default.pi} file and all response files for the
U, UVW1, UVM2, UVW2 filters were downloaded from the OM response file directory
in HEASARC Archive\footnote{http://heasarc.gsfc.nasa.gov/FTP/xmm/data/responses/om/}.
Each count rate and its associated error were entered into the default filter file and then
combined with the response file of the corresponding OM filter, again by using the {\sc GRPPHA} tool
to produce OM data that could be used in \xspec.

We were also able to carry out simultaneous optical observations from the Lijiang 2.4m for three of
the four RM-selected SEAMBHs (J075101, J081441 and J100055). The observations were done as close as possible
in time to the \xmm{} observations, as much as weather and telescope time limitations permitted (see
Table~\ref{table.data} for dates). Due to bad weather conditions J080101 was not observed. The observation
and the reduction of the optical spectra followd the standard technique that is described in \citet{Du2015}.
This procedure resulted with flux calibrated optical spectra.

The \xmm{} EPIC spectra are combined with the optical/UV photometry and
optical spectroscopy to construct a simultaneous broadband nuclear SED for each AGN. There is an
ubiquitous data gap in the far UV region due to absorption by Galactic gas and dust.
Since the intrinsic SED for our low-redshift, low-BH mass AGN peaks in this band, it is very difficult to estimate
the bolometric luminosity and hence the spin of the BH (see paper~{\sc I}).

\begin{table*}
  {\renewcommand{\arraystretch}{1.19}
  \caption{Information on the observed source and observations.}
  \label{table.data}
  \begin{tabular}{lcccccccrc}
    \toprule
    & & \multicolumn{3}{c}{RM results} & & \multicolumn{3}{c}{EPIC pn} & Lijiang \\\cmidrule{3-5}\cmidrule{7-9}
    $\qquad\,\,$Object & z & $\log{M_{BH}}$ & $\log{\dot{\mathcal{M}}}$ & $\log{L_{5100\mAA{}}}$ & $E(B-V)$ & date & $t$ & CR & date \\
    $\qquad$SDSS \dots & & $[M{\odot}]$ & & [erg s$^{-1}$]& Galactic &  &[ks] & [cts] & \\
    \midrule
    J075101.42+291419.1$^{\dagger}$ & 0.1208 & $7.16^{+0.17}_{-0.09}$ & $1.34^{+0.25}_{-0.41}$ & 44.48 & 0.042 & 2015-05-04 & 10.65 & 244   & 2015-05-04\\
    J080101.41+184840.7 & 0.1396 & $6.78^{+0.34}_{-0.17}$ & $2.33^{+0.39}_{-0.72}$ & 45.01 & 0.032 & 2015-10-10 & 21.35 & 38985 & $a$\\
    J081441.91+212918.5 & 0.1628 & $6.97^{+0.23}_{-0.27}$ & $1.56^{+0.63}_{-0.57}$ & 44.65 & 0.039 & 2015-10-15 & 12.22 & 23233 & 2015-10-19 \\
    J100055.71+314001.2$^{\ddagger}$ & 0.1948 & $6.50^{+0.20}_{-0.20}$ & $1.90^{+0.50}_{-0.50}$ & 44.12 & 0.017 & 2015-06-02 & 18.27 & 8477 & 2015-06-01\\
    \bottomrule
    \multicolumn{10}{l}{$^{\dagger}$The short exposure time does not allow to obtain good quality spectra above $\sim$3~keV.}\\
    \multicolumn{10}{l}{$^{\ddagger}$ The H$\beta$ lag was consistent with zero, so the BH mass and the accretion
    were estimated using the correlation between the BH}\\
    \multicolumn{10}{l}{$\,\,\,$ mass and $L_{5100}^{1/2}FWHM^2$, see Sect.~\ref{BHmassAccretionSection}.}\\
    \multicolumn{10}{l}{$^{a}$The object J080101 was not observed due to bad weather conditons.}\\
  \end{tabular}
}
\end{table*}

\subsection{Fitting thin AD models}
\label{fittingThinADmodels}
For the three RM-selected AGN with new simultaneous optical and X-ray observations (J075105, J081441 and J100055) we fitted thin AD models
to the new simultaneous observations following the methodology discussed in paper~{\sc I}.
We correct for Galactic extinction, subtracted the stellar and emission line contributions, and considered the possibility
of intrinsic reddening of the sources. For the Galactic interstellar reddening, we assumed the \citet{Cardelli1989}
extinction law using the Galactic extinction colour excess $E(B-V)$ obtained from the NASA/IPAC Infrared Science
Archieve\footnote{http://irsa,ipac.caltech.edu/applications/DUST}. The most consistent approach for
the intrinsic reddening is to add this as a free parameter in the disc modelling
analysis. The narrow spectral window of our data makes it impossible to distinguish between different extinction curves,
so the intrinsic reddening were modelled by a SMC extinction-like curve \citep{Gordon2003} that best represents typical
type-{\sc i} AGN. The host-galaxy contribution
was estimated using the fact that the broad H$\beta$ lines show no Baldwin effect \citep{Dietrich2002}.
This means that the observed equivalent width of the line, EW(H$\beta$), can be used to derive the fraction
of the non-AGN light entering the aperture at 4861\AA{}. We used an average EW(H$\beta$) of
$(89\pm31)$\AA{} which is representative of SEAMBHs \citep{DuPu2015}.

We used the \citet{Slone2012} code to calculate a large number of thin disc spectra that
include the entire range of BH mass and accretion rate expected in our sample. The fitting procedure includes the
comparison of the observed SED with various combination of disc SEDs covering the range of mass,
accretion rate and intrinsic reddening. The reddening is taken into account by changing $E(B-V)$
in steps of 0.004~mag, calculating, for each value, a new mass accretion rate and its range of uncertainty.
A simple $\chi^2$ procedure was used to find the best-fit combination of reddening and thin AD models. We use
three line-free windows covering the optical spectroscopic data, and the simultaneous
OM photometry. The line-free windows are centred on 4205\AA{}, 5100\AA{} and 6855\AA{}, with widths
ranging from 10 to 30\AA{}. For the error on each continuum point, we combine the standard error from the
Poison noise, an assumed 5\% error on the flux calibration, and the relative error of 20\% on the combination
of the uncertainties on the host-galaxy contribution and the unknown stellar population.
We refer the reader to paper~{\sc I} for a detailed description of the intrinsic SED recovering
and the optical/UV SED fitting analysis. We note that the disc spectrum
at the spectral window of $0.2-0.8\,\mu$m is insensitive to the spin given the assumed uncertainties
on the BH mass (0.5~dex) and the accretion rate (0.5~dex). Thus, the derived spin parameter
are not very meaningful and shorter wavelengths are required to estimate the spin.

\begin{figure*}
    \centering
    \includegraphics[width=0.32\linewidth]{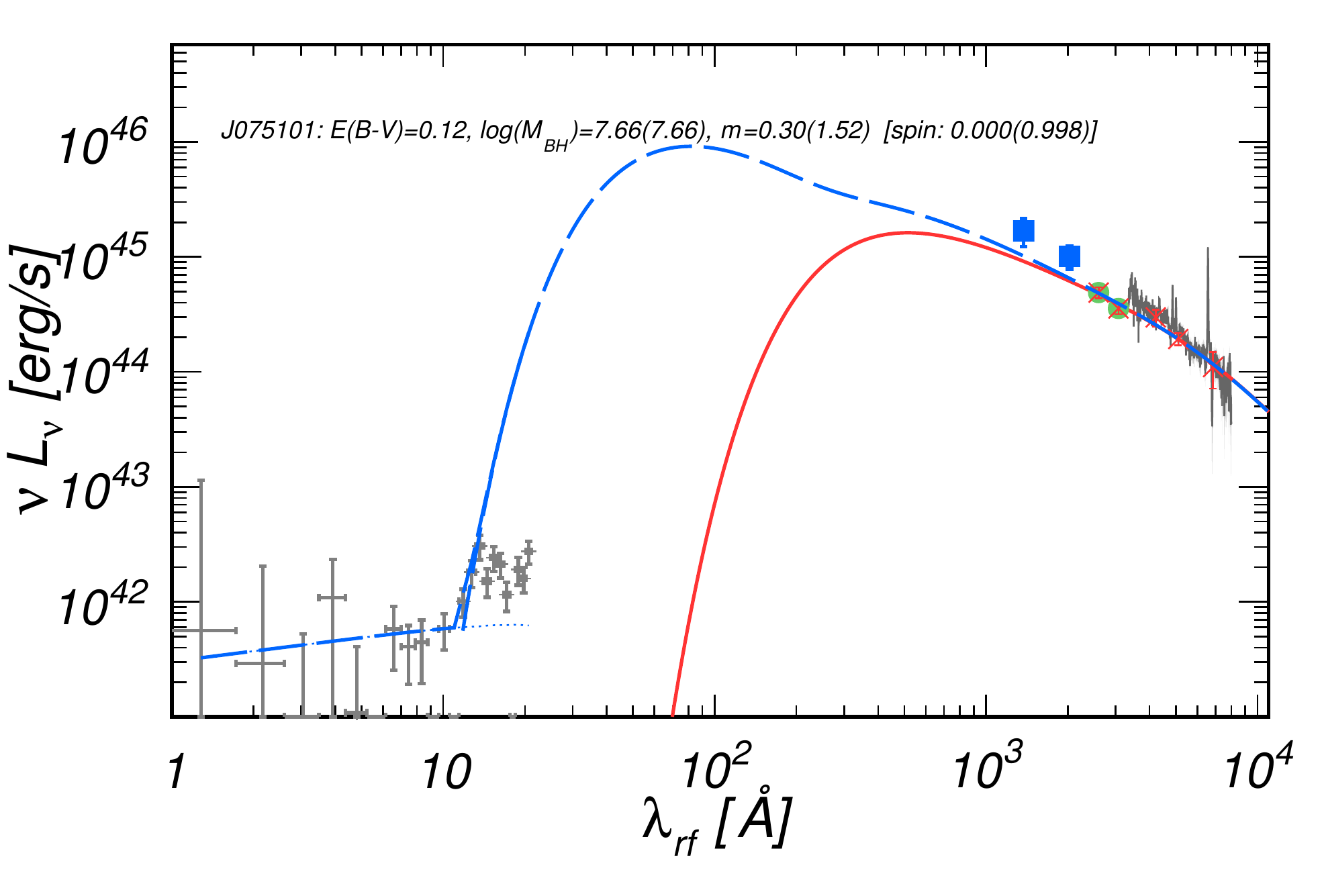}
    \includegraphics[width=0.32\linewidth]{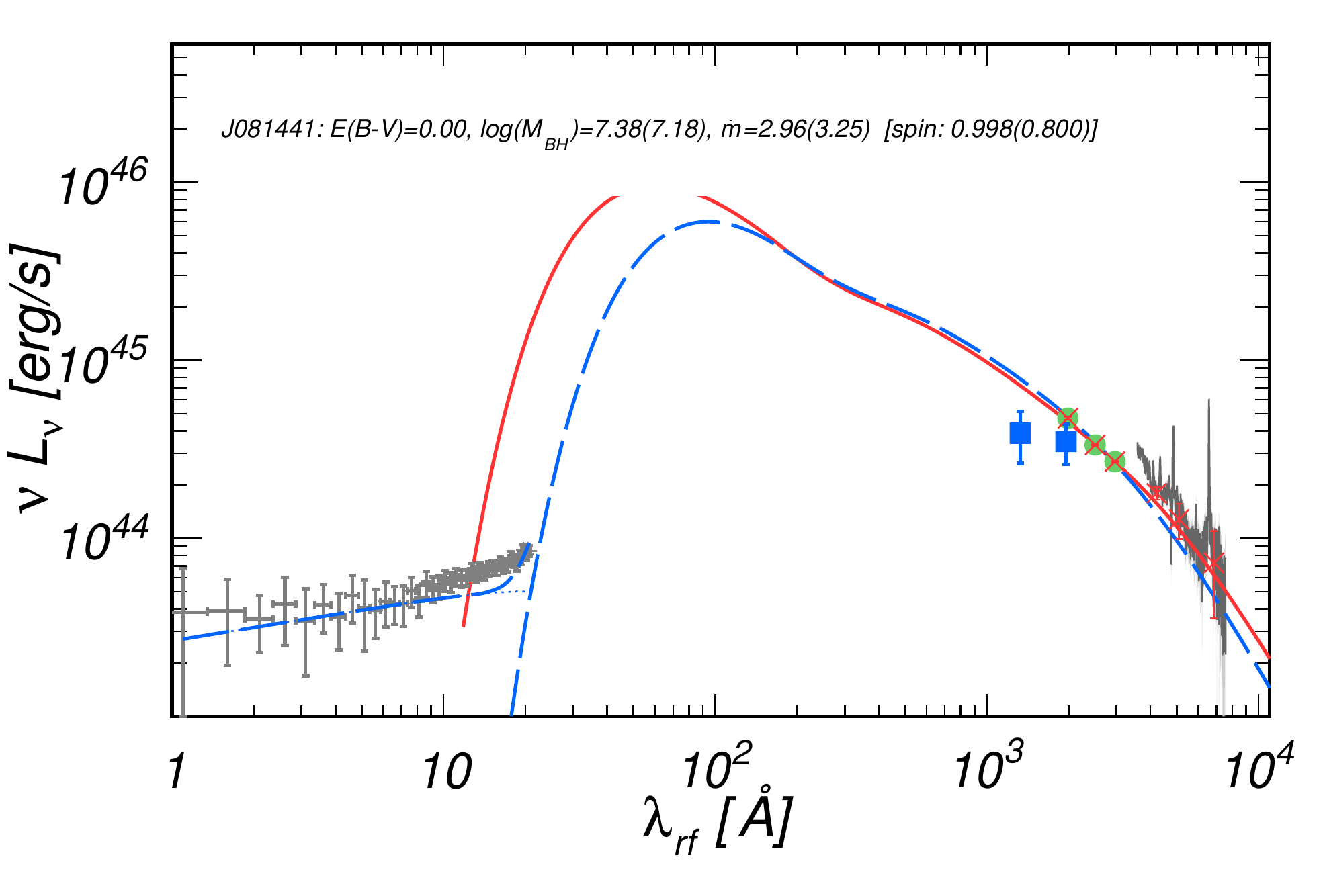}
    \includegraphics[width=0.32\linewidth]{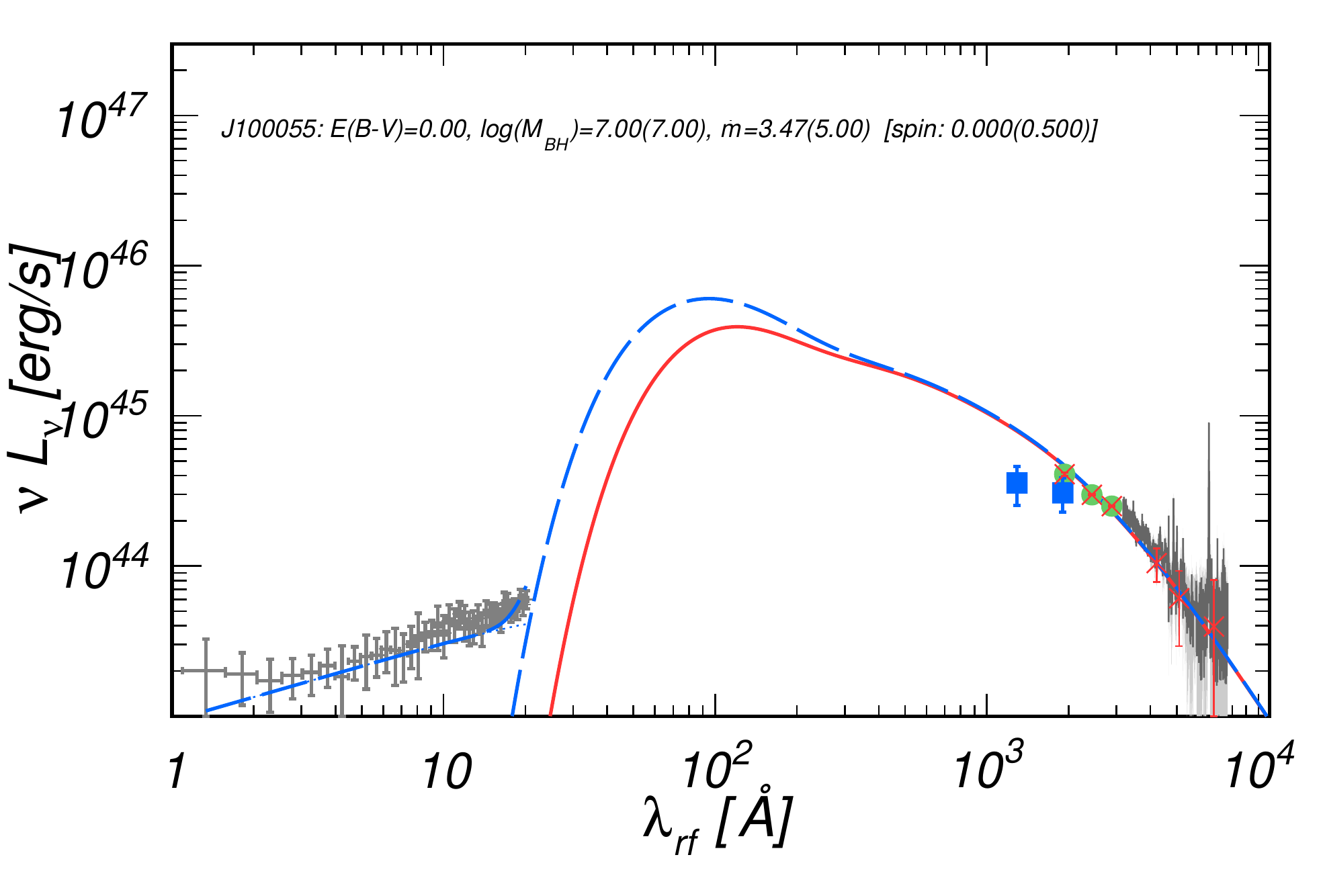}
    \caption{Simultaneous optical-to-X-ray SED with the best-fit thin AD model over-plotted assuming
    an inclination of $\cos{\theta}=0.75$. While the red line represents the best-fit thin AD models to the
    optical spectrum, the blue line represents the best fit to the optical-UV-X-ray SED assuming a power-law plus
    thin AD model as responsible of the hard-X-ray emission and the soft-X-ray-to-optical emission, respectively.
    Only the objects with simultaneous optical (green points: OM photometry; gray spectrum: Lijiang observations)
    and X-ray spectra are shown: J075101, J081441  and J100055, respectively.
    The blue square points represent the available non-contemporaneous UV photometry from GALEX which are not used
    in the fitting process.}\label{figure.broadbandSED}
\end{figure*}

Our best fitted AD SED are shown in Fig.~\ref{figure.broadbandSED} and all the model parameters, including
the disc luminosity, $L_{AGN}$, and the monochromatic luminosities at 2500~\AA{}, are listed in Table~\ref{table.discfit}.
For J075101, J080101 and J081441, a quiescent galaxy model from the evolutionary
spectral library of \citet{Charlot1991} with an age of 11~Gyr provided sufficiently
good fit to the stellar spectrum. On the basis of the "bluer" optical spectrum of J100055,
the stellar population must be younger than 11~Gyr. A template with 1.4~Gyr and solar metallicity
provides better fit to this spectrum and was adopted in this case.
As we expected from our previous work (paper~{\sc I}), the simultaneous optical/UV SED of the
SEAMBHs AGN, even those with the highest mass accretion rates, $\dot{\mathcal{M}}>20$, can be fit by the simple
optically thick, geometrically thin AD model. The declared good fits (see Fig.~\ref{figure.broadbandSED})
still include small deviations of the model from the local continuum at some wavelengths. This is not surprising given
the uncertainties on AD models, especially the radiative transfer in the disc atmosphere that was not treated
here in detail, as well as on the choice of the host galaxy template.

\begin{table*}
  {\renewcommand{\arraystretch}{1.19}
  \caption{Simultaneous optical/UV SED. Measured and deduced physical parameters while
            modelling the intrinsic reddening $E(B-V)$
            with a classical SMC extinction curve and assuming an inclination of $\cos{\theta}$=0.75.
            For J080101 there is no contemporaneous optical spectrum, and only two OM photometric points thus no model
            could be derived.}
  \label{table.discfit}
  \begin{tabular}{clcccccccccc}
    \toprule
    Object & \multicolumn{2}{c}{Host Galaxy}  && \multicolumn{5}{c}{thin AD model} &  \\\cmidrule{2-3}\cmidrule{5-11}
    & template & f &&  $E(B-V)$ & $\log{M_{BH}}$ & $\log{\dot{\mathcal{M}}}$ & $\eta$ & $\chi^{2}_{\nu}$ & $\log{L_{AGN}}$ & $\log{L_{2500\mAA}}$ &$\log{L_{2\text{keV}}}$ \\
        & & [\%] && intrinsic & $[M_{\odot}]$ & & & & [erg s$^{-1}$] & [erg s$^{-1}$] & [erg s$^{-1}$] \\
    \midrule
    J075101 & 11 Gyr & 0.26 && $0.120^{+0.240}_{-0.070}$ & 7.66 & 1.42 & 0.057 & 0.489 & 45.27 & 44.70 & 41.67  \\
    J081441 & 11 Gyr & 0.32 && $0.004^{+0.004}_{-0.004}$ & 7.28 & 1.38 & 0.321 & 2.803 & 45.41 & 44.54 & 43.67  \\
    J100055 & 1.4 Gyr& 0.70 && $0.000^{+0.010}_{-0.000}$ & 7.00 & 1.78 & 0.057 & 4.823 & 45.65 & 44.47 & 43.44  \\
    \bottomrule
  \end{tabular}
}
\end{table*}

\subsection{X-ray modelling}
\label{XrayAnalysis}
We performed an X-ray spectral analysis with \xspec{}~v12.8.2, using only the more accurately calibrated data
at 0.3--10$\,$keV. The source spectra for the three objects with the large number of counts were grouped to have
at least 20 counts in each bin in order to apply the modified~$\chi^{2}$ minimization technique. For the lowest
quality spectrum of J075101, with less than 400 counts, we used a minimum of 15 counts per bin instead of 20.
All quoted errors are for a 90 per cent confidence interval for one parameter ($\Delta\chi^2=2.706$).\\

We first consider the hard X-ray (2--10~keV) band, and apply a simple power law (PL) model (with neutral
absorption fixed at the Galactic value, {\sc wabs}) to all sources. In all the cases, we
obtain a satisfactory descriptions of the hard X-ray emission which allows us to constrain the spectral
photon index within an average error of 10\%. We can not constrain the spectral slope of J075101 due to the
poor signal-to-noise ratio above a few keV which results in a huge error on the slope and hence its optical-to-X-ray
spectral slope, $\alpha_{OX}$, can not be constrained. The extrapolation of the PL model to soft X-ray energies
does not provides an adequate description of the entire (0.3--10 keV) X-ray data which highlights an excess emission
to soft X-ray energies: the data below 2$\,$keV lie in all cases above the extrapolated bet-fit hard PL.
The hard X-ray spectra were also inspected for the presence of iron (Fe) emission lines. In all the objects
an additional narrow emission line in the range 6.4--6.7~keV (neutral to highly ionized Fe) does not improve the
significance of the fit. Because of this, we do not include an unresolved Fe K$\alpha$ emission line in our spectral
models.

In order to check the reliability of the soft excess, we refitted the whole 0.3--10~keV band with a simple PL
model and compare it with a (phenomenological) broken power law (BPL). Then we also explore the possibility of
X-ray absorption, however in no case the addition of a neutral absorption at the redshift of the source improved
the quality of the fit, so intrinsic absorption covering entirely the X-ray source plays a negligible role in
these RM-selected SEAMBHs. In Table~\ref{table.Xrayfit} we report the spectral slopes in the soft, $\Gamma_{s}$, and hard,
$\Gamma_{h}$, band for the broken PL model and the break energy which is in the 1.5--2~keV range.
The spectral slopes of the best-fit PL model over the entire X-ray band
were consistent with the soft slope of the BPL, $\Gamma_{s}$ which dominate the fit due to the high signal-to-noise
in the soft X-ray band. The BPL model is then a better description of the X-ray spectra of J080101, J081441 and
J100055, while a single PL model is adequate for J075101 (at the 90.36 per cent level only).
In other words, a soft excess is detected in J080101, J081441 and J100055, and not significantly present
in J075101 due to the poor signal-to-noise in the hard X-ray band.
The X-ray spectral slopes ($f_{\nu}\propto \nu^{\alpha_{X}}$) in the soft X-ray band, $\alpha_{X}$=1.62--1.96,
are consistent with the soft X-ray slope of NLS1, e.g. $\langle\alpha_X\rangle=$1.58 with a 16th and 84th
percentile of 1.11 and 2.16 found in the sample of \citet{Grupe2010} observed by {\it Swift XRT}.
The hard spectral slope is also remarkably similar ($\alpha_X$=1.10--1.27) to the typical spectral
slopes of NLS1 \citep[see e.g.][$\alpha_{X}\sim1.04$]{Zhou2011}. We conclude that the X-ray spectral shape of
our RM-selected SEAMBHs differ significantly from the average properties of low accreting AGN but
agree closely with the X-ray properties of NLS1s. We note that the broken PL is not entirely adequate
to reproduce the soft excess and that results
from such spectral fits should not be used to infer the significance of the soft excess detection.
A black body (BB) or a compton thick cloud (CC) representation are a better, though also phenomenological,
parametrization of the soft excess.

\begin{figure}
    \centering
    \includegraphics[width=0.42\linewidth]{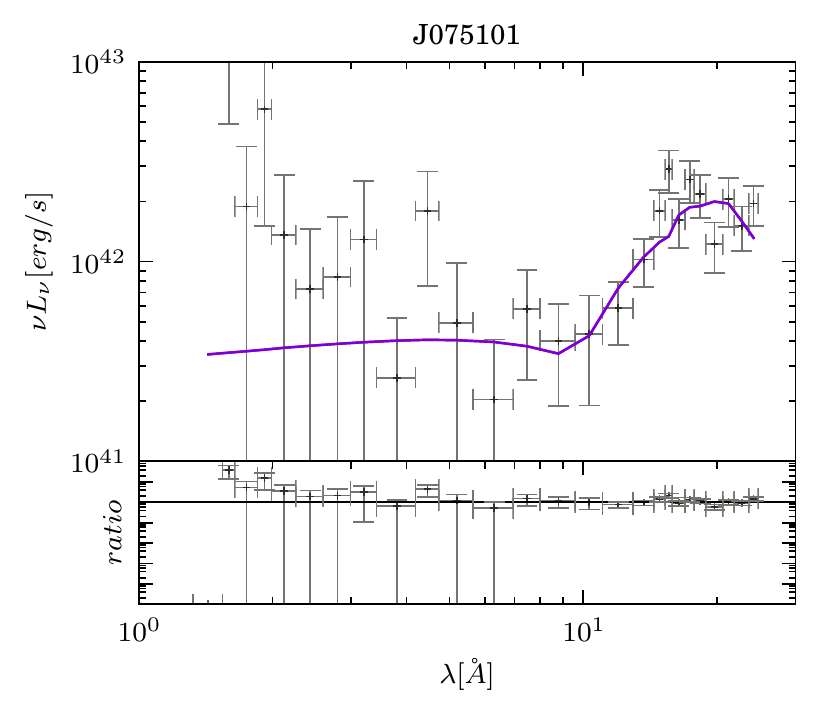}
    \includegraphics[width=0.42\linewidth]{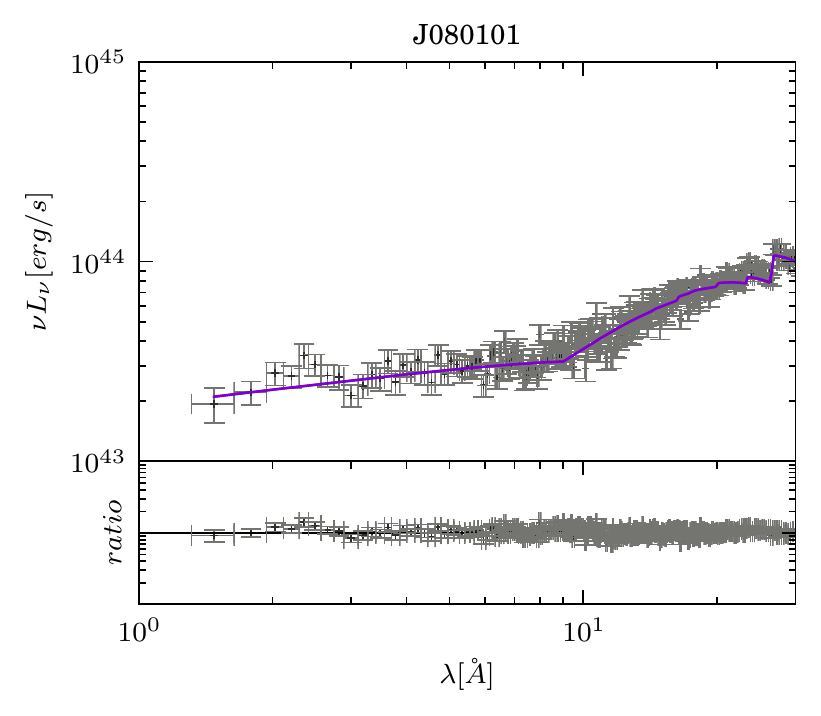}\\
    \includegraphics[width=0.42\linewidth]{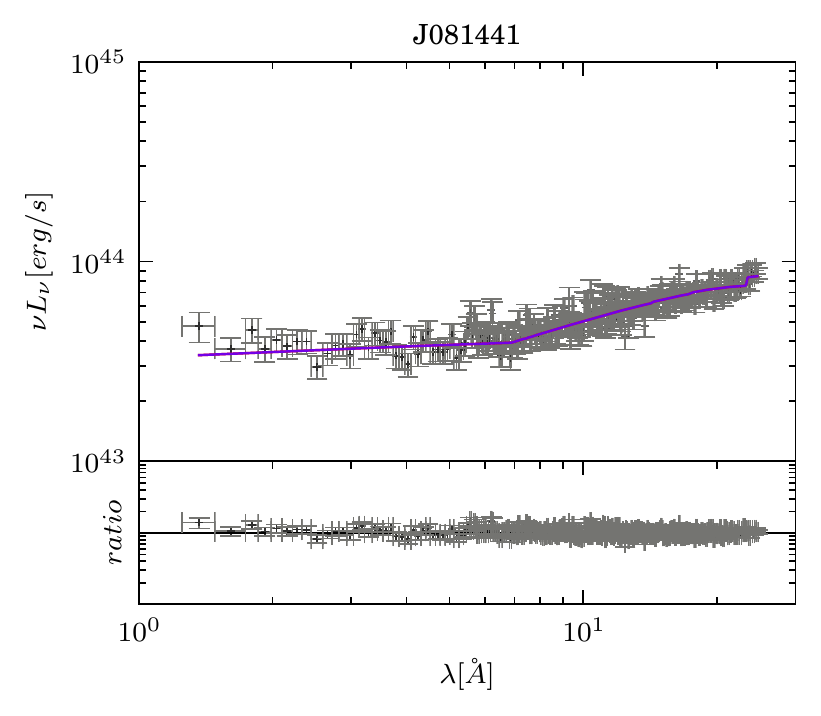}
    \includegraphics[width=0.42\linewidth]{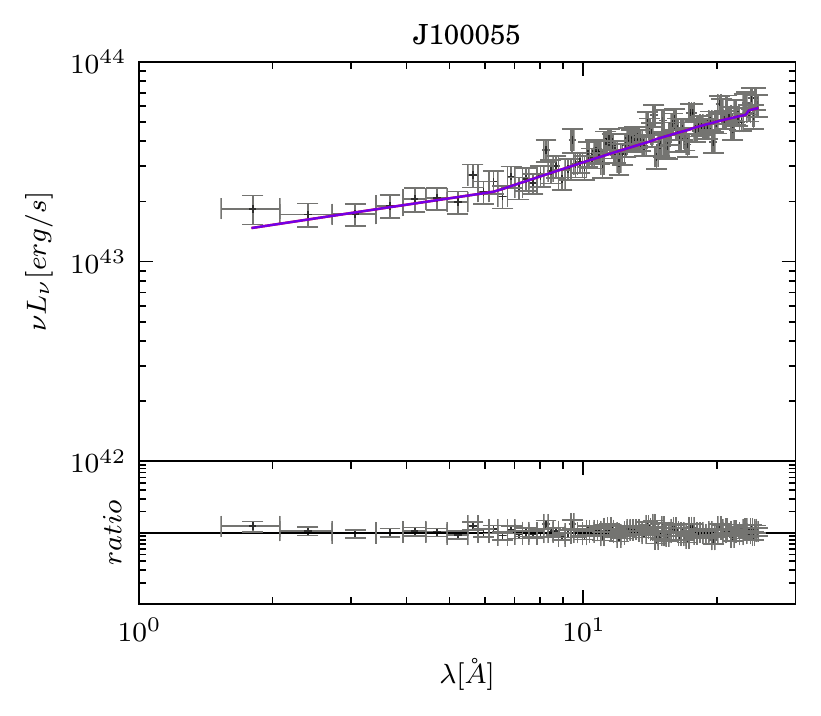}
    \caption{{\it XMM-Newton} spectra, best fitted broken-power law models and residuals for
        J075101, J080101, J081441 and J100055.
    The upper panels of each row show the results of the X-ray spectral fitting to the EPIC-pn spectra.
    The lower panels of each row show the data/model ratio of each data set.}\label{figure.Xrayfit}
\end{figure}

\begin{table*}
  {\renewcommand{\arraystretch}{1.19}
  \caption{Model parameters of the soft X-ray excess.
  Galactic absorption is always assumed ({\sc wabs}).
  Unabsorbed luminosities are given in units of $10^{43}$ erg$\,$s$^{-1}$ in the
  soft (0.5--2~keV) and hard (2--10~keV) energy bands. Following the work of \citet{Vasudevan2013},
  the term $L_{BB}/L_{PL}$ is defined as the ratio of the luminosity of
  the feature (using a black-body to model it) from 0.4--3 keV, to the luminosity
  in a relatively ``clean'', featureless portion of the primary power law from 1.5--6 keV.}
  \label{table.Xrayfit}
  \resizebox{0.99\linewidth}{!}{
  \begin{tabular}{clccccccccc}
    \toprule
      &    & Energy & \multicolumn{4}{c}{free paramaters} &  \\\cmidrule{4-7}
      Model$^{\dagger}$ & Object & Band  & $\Gamma_{s}$ & $E_{break}$ & $\Gamma_{h}$ & $kT_{e}$ & $\chi^{2}_{\nu}$ & $L_{BB}/L_{PL}^{\dagger}$ & \Lsoft & \Lhard \\
      &    & [keV]   & & [keV] &  & [keV] & & & $[10^{43}$ erg s$^{-1}$] & $[10^{43}$ erg s$^{-1}$]  \\
     \midrule
    PL&J075101 & 1.5--10   & - & - & $2.20^{+5.56}_{-2.08}$ & - & 1.541 & & 0.04\\
      &        & 0.5--10   & $3.42^{+0.21}_{-0.18}$ & - & - & - & 1.233 \\
      &J080101 & 2.0--10     & - & - & $2.09^{+0.12}_{-0.07}$ & - & 1.296 & & 3.60 \\
      &        & 0.5--10   & $2.90^{+0.02}_{-0.02}$ & - & - & - & 1.993 \\
      &J081441 & 2.0--10     & - & - & $2.08^{+0.07}_{-0.07}$ & & 1.160 & & 5.31\\
      &        & 0.5--10   & $2.56^{+0.03}_{-0.01}$ & - & - & - & 1.537 \\
      &J100055 & 2.0--10     & - & - & $2.27^{+0.17}_{-0.16}$ & - & 0.780 & & 2.82\\
      &        & 0.5--10   & $2.63^{+0.03}_{-0.05}$ & - & - & - & 1.040 \\\midrule
    BPL&J075101  & 0.5--10 & $2.62^{+0.60}_{-0.54}$ & $0.79^{+0.13}_{-0.11}$ & $2.2$ & - & 1.184 & & 3.68 & 0.04\\
        &J080101 & 0.5--10 & $2.96^{+0.02}_{-0.02}$ & $1.66^{+0.10}_{-0.13}$ & $2.23^{+0.30}_{-0.15}$ & - & 1.413 & & 10.96 & 3.61\\
        &J081441 & 0.5--10 & $2.62^{+0.02}_{-0.02}$ & $2.34^{+0.58}_{-0.70}$ & $1.97^{+0.18}_{-0.18}$ & - & 1.323 & & 9.29 & 5.33\\
        &J100055 & 0.5--10 & $2.68^{+0.04}_{-0.04}$ & $2.40^{+0.79}_{-0.70}$ & $2.31^{+0.23}_{-0.29}$ & - & 0.975 & & 2.04 & 0.97\\
    \midrule
    BB&J075101 & 0.5--10 & - & - & $2.2$                  & $0.132^{+0.013}_{-0.014}$ & 1.170 & $<4.49$ & 0.18 & 0.04  \\
      &J080101 & 0.5--10 & - & - & $2.27^{+0.04}_{-0.04}$ & $0.124^{+0.002}_{-0.002}$ & 1.292 & 1.39    & 7.94 & 3.55  \\
      &J081441 & 0.5--10 & - & - & $2.25^{+0.04}_{-0.04}$ & $0.132^{+0.004}_{-0.004}$ & 1.307 & 0.59    & 8.64 & 5.20  \\
      &J100055 & 0.5--10 & - & - & $2.52^{+0.07}_{-0.08}$ & $0.114^{+0.009}_{-0.010}$ & 0.992 & 0.47    & 2.03 & 0.91  \\
    \bottomrule
    \multicolumn{11}{l}{$^{\dagger}$ \xspec{} definition: {\sc PL)} \hardpl; {\sc BPL)} \bknpow; {\sc BB)}: \bb}\\
  \end{tabular}
}}
\end{table*}

We attempted to fit the entire optical-UV-X-ray data of J075101, J081441 and J100055 using their contemporaneous
optical/UV and X-ray observations that were presented here. The optical-UV part were modelled by the best-fit thin AD
model with the spin as the only free parameter which was combined with a single power-law
in order to fit simultaneously optical/UV and X-ray data sets. Our simultaneous best fits are shown in
Fig.~\ref{figure.broadbandSED}. Although the addition of the disc component to the X-ray emission improved the fit, our best-fit
continuum model significantly under-predicts the soft emission. This suggests that the AD can not explain the
total emission of the soft X-ray. In fact, SEAMBHs AD are predicted to be ``slim'' and their extreme UV
spectra are not fully understood. Because of the limited band of the X-ray data, and the large uncertainties on slim disc SEDs,
we are not in a position to resolve this issue by accurate spectral fitting of the data.
In fact, the SEAMBHs systems modelled by paper~{\sc I}, do not
show any (indirect) indications that their far UV luminosity is unusually high compared with the far-UV luminosity of sub-Eddington systems.
This might mean that this ``soft excess'' is not part of the big blue bump, however we think that the most likely explanation is
that SEAMBHs do not host thin disc, being in good agreement with the broad underlying assumption that SEAMBHs are the best candidates
to host slim disc geometries.

An additional consistency check can be obtained by studying the soft X-ray emission as a thermal emission of
a Compton thin plasma with certain temperature $kT_e$. In this case, we consider a simple continuum model
comprising a power-law plus black-body ({\sc zbbody}) emission to model the prominent soft excess. The addition
of this thermal component to the PL model improved the fit, providing an excellent fit in all the cases (see
table~\ref{table.Xrayfit} model BB). We find that the best-fit X-ray spectral index is also steep ($\alpha_X=1.2-1.5$) and consistent
within error to that obtained by fitting a simple power-law model to the hard X-ray band data.
The best-fit to this model for the source with no need for a soft excess (J075101) was also computed.
Finally, we parametrize the strength of the soft excess, $L_{BB}/L_{PL}$, following \citet{Vasudevan2013} by the ratio of
the luminosity of the feature (using a black-body to model it) from 0.4--3~keV, to the luminosity in a relatively ``clean'',
featureless portion of the primary power law from 1.5--6~keV. We find that the strength of the soft excess range from a factor 0.47 up to 1.4, and
despite the small size of the SEAMBHs, the average of the BB contribution in the 0.5--2~keV
clusters around a factor 0.8 being higher than previous studies \citep[][]{Vasudevan2013}.
In the case of J075101 (where there is no need for soft excess) the BB contribution is only an upper limit of
a factor 4.5. We find that the inferred temperature of the soft excess clusters with very little dispersion
around 114--143~eV with an average of 128~eV, while the disc luminosity of our sample spans about two orders
of magnitude. Although 128~eV is a reasonable value for the disc temperatures around our objects, it is
quite remarkable that the inferred temperature is very similar to that of PG quasars, casting doubts on its
interpretation in terms of thermal emission \citep{Vasudevan2013}.

\bsp    
\label{lastpage}
\end{document}